\documentclass[usenatbib]{mn2e}
\usepackage{times}
\usepackage{txfonts}
\usepackage{graphicx}
\usepackage{setspace}
\usepackage{natbib}
\usepackage{sidecap}
\usepackage{color}
\usepackage{subfigure}
\usepackage{floatflt}
\usepackage{rotating}
\usepackage{multirow}
\usepackage{aalongtable, lscape, afterpage}
\newcommand{\comment}[1]{}
\newcommand{\rul}{\rule[-2.50mm]{0mm}{2mm}}
\newcommand{\ffir}{F_{\st{FIR}}}

\begin{document}


\newcommand{\st}[1]{\mathrm{#1}} 
\newcommand{\pow}[2]{$\st{#1}^{#2}$}
\newcommand{\grad}{\hspace{-0.15em}\r{}}
\newcommand{\lm}{\lambda}
\newcommand{\average}[1]{\left\langle #1 \right\rangle}
\newcommand{\pp}[1]{#1^{\prime\prime}}
\newcommand{\p}[1]{#1^{\prime}}

\newcommand{\oi}{[O{\sc i}]}
\newcommand{\cii}{[C{\sc ii}]}
\newcommand{\nii}{[N{\sc ii}]}
\newcommand{\oiii}{[O{\sc iii}]}
\newcommand{\si}{[Si{\sc i}]}
\newcommand{\oib}{[O{\sc ib}]}
\newcommand{\mm}{~$\mu$m}
\newcommand{\ha}{H$\alpha$}
\newcommand{\civ}{[C{\sc iv}]}
\newcommand{\niiopt}{\nii~$\lambda~6583 \AA$}
\newcommand{\md}{M_{\st{d}}}
\newcommand{\td}{T_{\st{d}}}
\newcommand{\mdc}{M_{\st{d,c}}}
\newcommand{\tdc}{T_{\st{d,c}}}
\newcommand{\mdw}{M_{\st{d,w}}}
\newcommand{\tdw}{T_{\st{d,w}}}
\newcommand{\mg}{M_{\st{g}}}
\newcommand{\tcmb}{T_{\st{cmb}}}
\newcommand{\hextra}{H_{\st{extra}}}
\newcommand{\hcol}{N$_{\st H}$}
\newcommand{\pdv}{$P{\st d}V$}
\newcommand{\nitrogen}{$Z_{\odot}$(N)}

\newcommand{\h}{~h_{71}~}
\newcommand{\hinv}[1]{~h_{71}^{#1}~}
\newcommand{\eq}[1]{(\ref{eq-#1})}
\newcommand{\wrt}{with respect to\ }
\newcommand{\mms}{\frac{M_{\odot}}{M}}
\newcommand{\mpy}{\ms~\st{yr}^{-1}}
\newcommand{\ct}{t_{\st{cool}}}
\newcommand{\rc}{r_{\st{cool}}}
\newcommand{\lc}{L_{\st{cool}}}
\newcommand{\tvir}{T_{\st{vir}}}
\newcommand{\rvir}{R_{\st{500}}}
\newcommand{\mvir}{M_{\st{500}}}
\newcommand{\mbh}{M_{\st{BH}}}
\newcommand{\ms}{M_{\odot}}
\newcommand{\ls}{L_{\odot}}
\newcommand{\lxb}{L_{\st {Xb}}}
\newcommand{\lfir}{L_{\st{FIR}}}
\newcommand{\lfirtot}{L_{\st{FIR,tot}}}
\newcommand{\lx}{L_{\st {X}}}
\newcommand{\lr}{L_{\st {R}}}
\newcommand{\lbcg}{L_{\st{BCG}}}
\newcommand{\lt}{L_{\st{X}}{\st -}\tvir}
\newcommand{\Dl}{D_{\st{L}}}
\newcommand{\Da}{D_{\st{A}}}
\newcommand{\mdr}{\dot{M}_{\st{classical}}}
\newcommand{\smdr}{\dot{M}_{\st{spec}}}
\newcommand{\ohb}{[\st{O_{~III}}]~\lambda 5007/{\st H}\beta}
\newcommand{\norm}{$\eta_{\st{OSP}}$}

\newcommand{\chandra}{\textit{Chandra}}
\newcommand{\vla}{\textit{VLA}}
\newcommand{\gmrt}{\textit{GMRT}}
\newcommand{\atca}{\textit{ATCA}}
\newcommand{\XMM}{\textit{XMM-Newton}}
\newcommand{\einstein}{\textit{Einstein}}
\newcommand{\asca}{\textit{ASCA}}
\newcommand{\rosat}{\textit{ROSAT}}
\newcommand{\herschel}{\textit{Herschel}}
\newcommand{\iras}{\textit{IRAS}}
\newcommand{\spitzer}{\textit{Spitzer}}
\newcommand{\hiflux}{\textit{HIFLUGCS}}

%
\def\aj{AJ}%
\def\araa{ARA\&A}%
\def\apj{ApJ}%
\def\apjl{ApJ}%
\def\apjs{ApJS}%
\def\ao{Appl.~Opt.}%
\def\apss{Ap\&SS}%
\def\aap{A\&A}%
\def\aapr{A\&A~Rev.}%
\def\aaps{A\&AS}%
\def\azh{AZh}%
\def\baas{BAAS}%
\def\jrasc{JRASC}%
\def\memras{MmRAS}%
\def\mnras{MNRAS}%
\def\pra{Phys.~Rev.~A}%
\def\prb{Phys.~Rev.~B}%
\def\prc{Phys.~Rev.~C}%
\def\prd{Phys.~Rev.~D}%
\def\pre{Phys.~Rev.~E}%
\def\prl{Phys.~Rev.~Lett.}%
\def\pasp{PASP}%
\def\pasj{PASJ}%
\def\qjras{QJRAS}%
\def\skytel{S\&T}%
\def\solphys{Sol.~Phys.}%
\def\sovast{Soviet~Ast.}%
\def\ssr{Space~Sci.~Rev.}%
\def\zap{ZAp}%
\def\nat{Nature}%
\def\iaucirc{IAU~Circ.}%
\def\aplett{Astrophys.~Lett.}%
\def\apspr{Astrophys.~Space~Phys.~Res.}%
\def\bain{Bull.~Astron.~Inst.~Netherlands}%
\def\fcp{Fund.~Cosmic~Phys.}%
\def\gca{Geochim.~Cosmochim.~Acta}%
\def\grl{Geophys.~Res.~Lett.}%
\def\jcp{J.~Chem.~Phys.}%
\def\jgr{J.~Geophys.~Res.}%
\def\jqsrt{J.~Quant.~Spec.~Radiat.~Transf.}%
\def\memsai{Mem.~Soc.~Astron.~Italiana}%
\def\nphysa{Nucl.~Phys.~A}%
\def\physrep{Phys.~Rep.}%
\def\physscr{Phys.~Scr}%
\def\planss{Planet.~Space~Sci.}%
\def\procspie{Proc.~SPIE}%

\title[Herschel Observations of the Centaurus
Cluster]{Herschel\thanks{{\it Herschel} is an ESA space observatory
    with science instruments provided by European-led Principal
    Investigator consortia and with important participation from
    NASA.} observations of the Centaurus cluster - the dynamics of
  cold gas in a cool core}

\author[Rupal Mittal et al.]{
\centerline{R. Mittal,$^{1}$ 
C. P. O'Dea,$^{2}$ 
G. Ferland,$^{3}$ 
J. B. R. Oonk,$^{4,5}$ 
A. C. Edge,$^{6}$
R.~E.~A.~Canning,$^{7}$}  
\newauthor 
\centerline{H. Russell,$^{8}$ 
S. A. Baum,$^{1}$ 
H. B\"ohringer,$^{9}$ 
F. Combes,$^{10}$
M. Donahue,$^{11}$
A. C. Fabian,$^{7}$} 
\newauthor 
\centerline{N. A. Hatch,$^{12}$
A. Hoffer,$^{11}$
R. Johnstone,$^{7}$
B. R. McNamara,$^{8,13,14}$} 
\newauthor 
\centerline{P. Salom\'e,$^{10}$
and G. Tremblay$^{15}$}\\\\
$^{1}$  Chester F. Carlson Center for Imaging Science, Rochester Institute of Technology, Rochester, NY 14623, USA \\ 
$^{2}$  Department of Physics, Rochester Institute of Technology, 84 Lomb Memorial Drive, Rochester, NY 14623, USA \\
$^{3}$  Department of Physics, University of Kentucky, Lexington, KY 40506, USA  \\
$^{4}$  Leiden Observatory, Leiden University, P.B. 9513, Leiden 2300 RA, The Netherlands \\
$^{5}$  Netherlands Institute for Radio Astronomy, Postbus 2, 7990 AA Dwingeloo, The Netherlands \\
$^{6}$  Institute for Computational Cosmology, Department of Physics, Durham University, Durham, DH1 3LE, UK \\
$^{7}$  Institute of Astronomy, Madingley Road, Cambridge, CB3 0HA, UK \\
$^{8}$  Department of Physics \& Astronomy, University of Waterloo, Canada, N2L 3G1 \\
$^{9}$  Max-Planck-Institut f\"ur extraterrestrische Physik, 85748 Garching, Germany \\
$^{10}$  Observatoire de Paris, LERMA, CNRS, 61 Av. de l'Observatoire, 75014 Paris, France \\
$^{11}$  Michigan State University, Physics and Astronomy Dept., East Lansing, MI 48824, USA \\
$^{12}$  School of Physics and Astronomy, University of Nottingham, University Park, Nottingham NG7 2RD, UK \\
$^{13}$  Perimeter Institute for Theoretical Physics, Waterloo, Canada \\
$^{14}$  Harvard-Smithsonian Center for Astrophysics, 60 Garden Street, Cambridge, MA, USA \\
$^{15}$  Astrophysical Science and Technology, Rochester Institute of Technology, Rochester, NY 14623, USA
}

\date{Received/Accepted}

\maketitle

\begin{abstract}

  Brightest cluster galaxies~(BCGs) in the cores of galaxy
  clusters have distinctly different properties from other low
  redshift massive ellipticals. The majority of the BCGs in cool-core
  clusters show signs of active star formation. We present
  observations of NGC~4696, the BCG of the Centaurus galaxy cluster,
  at far-infrared~(FIR) wavelengths with the {\herschel} space
  telescope. Using the PACS spectrometer, we detect the two strongest
  coolants of the interstellar medium, {\cii} at 157.74{\mm} and {\oi}
  at 63.18{\mm}, and in addition {\nii} at 121.90{\mm}. The {\cii}
  emission is extended over a region of 7~kpc with a similar spatial
  morphology and kinematics to the optical {\ha} emission. This has
  the profound implication that the optical hydrogen recombination
  line, {\ha}, the optical forbidden lines, {\niiopt}, the soft X-ray
  filaments and the far-infrared {\cii} line all have the same energy
  source.

  We also detect dust emission using the PACS and SPIRE photometers at
  all six wavebands. We perform a detailed spectral energy
  distribution fitting using a two-component modified black-body
  function and find a cold 19~K dust component with mass
  1.6$\times10^6~\ms$ and a warm 46~K dust component with mass
  4.0$\times10^3~\ms$. The total FIR luminosity between 8{\mm} and
  1000{\mm} is 7.5$\times10^8~\ls$, which using Kennicutt relation
  yields a low star formation rate of 0.13~$\mpy$. This value is
  consistent with values derived from other tracers, such as
  ultraviolet emission. Combining the spectroscopic and photometric
  results together with optical {\ha}, we model emitting clouds
  consisting of photodissociation regions~(PDRs) adjacent to ionized
  regions. We show that in addition to old and young stellar
  populations, there is another source of energy, such as cosmic rays,
  shocks or reconnection diffusion, required to excite the {\ha} and
  {\cii} filaments.
\end{abstract}

\newcommand{\cen}{NGC~4696}

\section{Introduction}
\label{intro}

Clusters of galaxies offer us a unique opportunity to study
astrophysical components on widely differing scales. Intensive
theoretical and observational efforts have revealed that these
components are closely tied to one another. The megaparsec-scale
intracluster medium~(ICM) is a hot plasma emitting bremsstrahlung
X-ray radiation. The central regions of this plasma ($\leq 300~$kpc)
have high electron densities; hence the numerous observed clusters
with peaked surface-brightness profiles. Such high gas densities imply
rapid gas cooling ($\ll 1/\st{H}_0$) leading to a steady, mass
deposition rate (tens to a few hundreds $\ms$~yr$^{-1}$)
\citep[e.g.][]{Fabian1994} and iteratively increasing central
densities.

However, this picture is far from complete. High-resolution $\XMM$
spectral and $\chandra$ imaging of cool-core~(CC) clusters
\cite[e.g.,][]{Tamura2001,Peterson2001,Kaastra2001a,Peterson2003,Xu2002,Sakelliou2002,Sanders2008,Hudson2010}
have shown very little cool gas in their cores with the central
temperatures no lower than a fifth of the virial temperature. Heating
by clustercentric active galactic nuclei~(AGN) through outflows
(possibly in conjunction with other processes like heat conduction,
cosmic-ray heating and convection) appears to be an irreplaceable
heating mechanism that prevents the gas from undergoing excessive
cooling
\citep[e.g.][]{Binney1995,Churazov2002,Roychowdhury2004,Voit2005}.
Several studies have established morphological, statistical and
physical correlations between X-ray properties of CC clusters and AGN
at their centers
\citep{Burns1990,Birzan2004,Rafferty2006,Peterson2006,Mittal2009,McNamara2007,Dunn2006,Edwards2007}
that support active galactic nuclei as the primary heating candidates.

Despite AGN and other plausible sources of heating which stop the ICM
from cooling catastrophically, recent observations of many ($\sim 40$)
brightest cluster galaxies (BCGs) show that the intracluster medium
gas {\it is} cooling but at a much suppressed level. They appear to
have substantial amount of cold gas, of which a small fraction is
forming stars
\cite[e.g.,][]{Johnstone1987,McNamara1989,Edge2003,ODea2004,ODea2010,ODea2008,Hicks2005,Mittaz2001,Allen1995}.
It seems that the cold gas is able to somehow survive for long periods
of time.

A significant number of cool-core clusters show optical
line-filaments, such as in A~426 \citep[Perseus,][]{Conselice2001},
A~1795 \citep{Crawford2005b}. \cite{Crawford1999} did an optical study
of $> 250$ dominant galaxies in X-ray selected galaxy clusters and
showed that about 25~\% of them have {\ha} emission-line in their
spectra with intensity ratios typical of cooling flow nebulae
\cite[also see][]{Hu1985,Heckman1989}.  Amidst these exciting
discoveries of {\ha}, \cite{Canning2010} have detected the optical
coronal line emission [FeX]~$\lambda$~6374$\AA$ in {\cen} using VIMOS
spectra, implying yet another component of gas at temperatures in the
range $(1$ to $5)\times10^6$~K \citep[also
see][]{Oegerle2001,Bregman2006}. Warm molecular H$_2$ at $\sim 2000$~K
\citep{Jaffe1997,Donahue2000,Edge2002,Hatch2005}, cold molecular H$_2$
at (300-400)~K \citep{Johnstone2007} and cold CO gas at few tens of
kelvin \citep{Edge2001,Salome2008} also exists in the cores and also
regions overlapping with the extended optical
filaments. \cite{Edge2001} detected 16 CC clusters with IRAM and JCMT,
implying a substantial mass $(10^{9-11.5})~\ms$ of molecular gas
within 50~kpc radius of the BCGs. Similarly, \cite{Salome2003} studied
32 BCGs in cool-core clusters with the IRAM~30~m telescope and found
gas masses between $3\times 10^8~\ms$ to $4\times10^{10}~\ms$.

From the above observations of CC clusters it is clear that there are
different components of gas at almost every temperature in the range
varying from the virial temperature of the ICM ($10^7$~K) to the
temperature of the star-forming molecular clouds ($10$~K). An
important result from the point of view of energetics of the observed
warm molecular hydrogen emission and also the optical coronal emission
line is that these emissions are very likely due to the gas being
reheated rather than cooling out of the ICM
\citep{Donahue2000,Canning2010}.  The molecular H$_2$ emission, for
example, is too bright to arise from gas cooling out of the ICM. Based
on the measured H$ _2$ line luminosity and assuming the fraction of
cooling in the line emission, $\eta=2~\%-10~\%$, the estimated mass
cooling rate is two orders of magnitude larger than the mass
deposition rate predicted from the X-ray surface brightness
profile. Understanding the details of how the mass and energy transfer
occurs is crucial. To that end a key component is the molecular gas
and dust at $< 60$~K, whose natural emission is accessible by
{\herschel}.

In this paper, we study NGC~4696, the brightest galaxy of the
Centaurus cluster of galaxies~(Abell~3526). This work is a part of a
Herschel Open Time Key Project devoted to study cold gas and dust in
11 BCGs. Preliminary results of this study can be found in
\cite{Edge2010a} and \cite{Edge2010b}.  {\cen} is at a redshift of
0.01016 \citep{Postman1995} corresponding to a radial velocity of
3045~km~s$^{-1}$. Owing to its proximity, this cluster has been a
subject of numerous studies
\citep[e.g.][]{Fabian1982,Lucey1986,deJong1990,Sparks1989,ODea1994,Allen1994,Sparks1997,Laine2003,Crawford2005,Taylor2006,Johnstone2007,Canning2010,Farage2010}. The
$\chandra$ X-ray observations show a bright dense core with plume-like
structures that spiral off clockwise to the northeast. It is a classic
cool-core cluster based on its short central [at 5~kpc
($0.004~\rvir$\footnote{$\rvir$ is the radius within which the average
  mass density of the cluster is 500 times the critical density of the
  Universe.})] cooling time ($< 0.5~$Gyr). However, the expected gas
mass deposition rate within the cooling region is relatively small,
$\sim$ a few tens of solar masses per year
\citep{Fabian1982,Hudson2010,Ikebe1999,Sanders2002}. Detailed {\XMM}
Reflection Grating Spectrometer~(RGS) observations indicate the
presence of cool gas at the center of the cluster with temperature in
the range of (0.35 to 3.7)~keV. Optical observations reveal bright
line-emitting filaments in {\niiopt} and {\ha}, which were
first discovered by \cite{Fabian1982} and have been mapped more
recently using the EMMI instrument on the 3.58~m New Technology
Telescope with a much higher resolution by \cite{Crawford2005}. These
filaments extend out in a similar manner to the spiral structure seen
in the X-ray. They also show a remarkable spatial correlation with the
dust features, in particular the dust lane seen looping around the
core of the {\cen} \citep{Sparks1989,Laine2003}.

{\cen} is a host to the low-power steep-spectrum FR~I radio source,
PKS~1246-410 \citep[e.g.][]{Taylor2006}. The total radio luminosity of
the radio source is $9.43\times 10^{40}~$erg~s$^{-1}$
\citep{Mittal2009}. The radio core coincides well with the duo-core
optical nucleus revealed by the high-resolution HST imaging. The radio
emission, which shows a one-sided jet oriented to the south on small
($<30$pc) scales that into lobes oriented east-west on kpc-scales
\citep{Taylor2006}.  The total extent of the lobes along the east-west
direction is about 10~kpc after which both the lobes eventually turn
south. The radio plasma is clearly interacting with the hot X-ray gas
and seems to be correlated with the X-ray cavities.

We describe the Herschel observations and data reduction in
Sect.~\ref{hobs} and the results in Sect.~\ref{results}. We present an
analysis of the AGN contribution and kinematics in
Sect.~\ref{analysis} and a detailed modeling of the photodissociation
region in Sect.~\ref{pdr}. We give our final conclusions in
Sect.~\ref{conclusions} and a summary of this study in
Sect.~\ref{summary}.  We assume throughout this paper the $\Lambda$CDM
concordance Universe, with $H_0 = 71~h_{71}$~km~s$^{-1}$~Mpc$^{-1}$,
$\Omega_{\st m} = 0.27$ and $\Omega_{\Lambda} = 0.73$.  This
translates into a physical scale of $\pp{1}=0.2$~kpc at the redshift
of {\cen}.

\section{Herschel Observations and Data Analysis}
\label{hobs}

We used the ESA Herschel space observatory \citep{Pilbratt2010} using
PACS \citep{Poglitsch2010} and SPIRE \citep{Griffin2010} to study the
cold phase ($<60$~K) of the interstellar and intracluster medium. The
aim of these observations was to understand the details of mass and
energy transfer between the different phases of gas. In particular, we
study photodissociation regions~(PDRs) which are relatively small in
volume-filling factor but owing to high densities (and thermal
pressures) as compared to that of the average interstellar
medium~(ISM) dominate the radiation of a galaxy
\citep[e.g.][]{Hollenbach1999}. PDRs define the emission
characteristics of the ISM and star formation regions in a
galaxy. They primarily comprise molecular hydrogen, ionized carbon,
neutral oxygen, CO and dust, such as silicates, silicate carbides,
poly aromatic hydrocarbons~(PAHs) etc. Detailed PDR descriptions and
chemical processes can be found in \cite[e.g.][]{Hollenbach1999,
  Kaufman1999,Rollig2007}.

\subsection{PACS Spectrometry}
\label{pacspec}

\begin{table*}
  \centering
  \caption{\small {\herschel} PACS spectroscopy observational log of {\cen} at a redshift of 0.01016. 
    All the lines were observed in the line spectroscopy mode and on the same day: 
    30$^{\st{th}}$~Dec.~2009.}
  \label{obs}
  \begin{tabular}{|c| c| c| c| c c| c c| c| c |}
  \hline
  Line &  Peak Rest $\lambda$ & ObsID & Duration & \multicolumn{2}{c|}{Bandwidth} & \multicolumn{2}{c|}{Spectral FWHM} & Spatial FWHM & Mode \\
         &  ({\mm})                       &            &      (s)      & ({\mm}) & (km~s$^{-1}$)       & ({\mm}) & (km~s$^{-1}$) &  & \\
  \hline\hline
  OI   &  63.180   & 1342188700  & 6912 & 0.266 & 1250       & 0.017    & 79  & $\pp{4.6}$  & 3x3 raster, step size $\pp{23.5}$\\
  CII  &  157.74   & 1342188700  & 3096 & 1.499 & 2820       & 0.126    & 237 & $\pp{11.5}$   & 3x3 raster, step size $\pp{23.5}$   \\
  NII  & 121.90    & 1342188701  & 3440 & 1.717 & 4180       & 0.116    & 280 & $\pp{8.9}$   & pointed   \\   
  OIII & 88.36      & 1342188701  & 3680 & 0.495 & 1660       & 0.033    & 110 & $\pp{6.4}$   & pointed   \\ 
  OIb & 145.52    & 1342188702  & 3440 & 1.576 & 3215       & 0.123    & 250 & $\pp{10.5}$   & pointed   \\   
  SiI   & 68.47      & 1342188702  & 3840 & 0.218 & 945         & 0.014    & 62   & $\pp{4.9}$   & pointed   \\
  \hline
  \end{tabular}
\end{table*}

We observed the two primary coolants of the neutral ISM, the {\cii}
line at 157.74{\mm} and the {\oi} line at 63.18{\mm}, along with
{\oib} at 145.52{\mm}, {\si} at 68.470{\mm}, {\nii} at 121.90{\mm} and
{\oiii} at 88.36{\mm} with the PACS spectrometer on {\it Herschel}.
The $^2$P$_{3/2} \rightarrow ^2$P$_{1/2}$ fine-structure emission line
of {\cii} is very often the brightest emission line in galaxy spectra,
followed by the {\nii} lines at 122{\mm} ($^3$P$_2 \rightarrow
^3$P$_1$) and 205{\mm} ($^3$P$_1 \rightarrow ^3$P$_0$). Ionized
nitrogen and oxygen are mainly produced in warm ionized medium, such
as H{\sc ii} regions, and so the {\nii} 122{\mm} and the {\oiii}
88.36{\mm} lines can be used to obtain constraints on the fractions of
ionized and neutral media.

All lines were observed in the line-spectroscopy mode with the
chopping and nodding implementation to subtract the rapidly varying
telescope background and dark current. The details of the
observational parameters are summarized in table~\ref{obs}. We used
the large chopper throw throughout, such that the mirror chops two
regions of sky $\p{6}$ apart alternately. While {\oi} and {\cii}
observations were made in the raster-mapping mode, all the other line
observations were made in the pointed mode with a field of view of
$\pp{47} \times \pp{47}$. The PACS spectrometer contains an image
slicer unit for integral field spectroscopy.  The slicer transforms a
5x5 pixel focal plane image into a linear array of 25 spatial pixels
termed $spaxels$, each $\pp{9.4} \times \pp{9.4}$ in size. This
technique ensures high sensitivity to detecting weak emission lines
and provides simultaneous spectra of all extended emission in the
field of view. The signal from each spaxel then goes through the
grating assembly resulting in 16 spectral elements. The PACS
spectrometer has two channels, red (51-105){\mm} and blue
(102-220){\mm}, which can be operated at the same time. This allowed
us to perform sequential observations of both {\oi} and {\cii}. The
raster mapping consisted of a 3 by 3 array of overlapping single
pointings (3 raster lines and 3 points per line) separated by
$\pp{23.5}$ in both the directions, along and orthogonal to the raster
lines.

We applied the standard pipeline routines described in the PACS data
reduction guideline~(PDRG) to process the data from their raw state
(level 0) to a fully-calibrated state (level 2), using the Herschel
Interactive Processing Environment~(HIPE) \citep{Ott2010} version
3.0~CIB~1475.  The data-reduction steps included removing cosmic-ray
glitches and correcting for the intrinsic non-linearities. We used the
ground-based flat-field estimates to correct for the pixel responses,
which are known to overestimate the fluxes. We followed the PACS
spectroscopy performance and calibration~(PSPC) memo to correct the
fluxes by applying factors of 1.3 for the blue band and 1.1 for the
red band.  The penultimate final PACS data product is in the form of
two $5 \times 5 \times n$ cubes, corresponding to the two nodding
positions, where $n$ is the number of wavelengths present. This
product is further rebinned in wavelength in accordance with the
actual wavelengths present and the user-requested resolution (see
Sect.~\ref{specanalysis1}). The PACS spatial resolution varies from
$\pp{11.3}$ for the {\cii} 157.74{\mm} line to $\pp{4.5}$ for the
{\oi} 63.18{\mm} line.

\subsubsection{Over- and up-sampling factors}
\label{specanalysis1}

As mentioned above, the $5 \times 5 \times n$ cubes are further
rebinned in wavelength by applying two factors called `oversample' and
`upsample'. The oversample factor increases the number of bins over
which the data are averaged to create a spectrum and the upsample
factor specifies the shift in wavelength in the units of the
binwidth. The default values, oversample=2 and upsample=1, correspond
to the signal sampled at the Nyquist-Shannon rate, yielding the
instrument spectral resolution at that wavelength.  Increasing the
upsample factor, for example, to upsample=2 yields two spectra, both
adhering to the instrument resolution but shifted by half a binwidth
with respect to each other. An upsample factor greater than unity may
be used to identify features narrower than the instrument resolution
in the line profile but at the expense of increased noise. Lastly, the
combination oversample=1 and upsample=1 undersamples the data to
render a spectral resolution a factor of two coarser than the original
one, hence affecting the full-width half maximum~(FWHM) of the fitted
line. We investigated various combinations of oversample and upsample
factors for each of the lines to optimize the signal to noise ratio
and the spectral resolution.

\subsubsection{Line Flux Estimation}
\label{specanalysis2}

In the case of emission well-centered on the central spaxel, a
wavelength-dependent point-source correction factor has to be applied
to the measured line flux-density to account for the fraction of the
beam that is lost to the neighbouring spaxels~(beam spillover). An
alternate way of recovering the flux in the entire beam is to co-add
the spectra obtained in different spaxels but this method suffers from
the risk of line distortion.  A yet alternate method is to keep the
spectra separate and add only the fluxes contained in the spaxels with
significant point source contribution. This method works well also for
slightly extended sources provided that the emission is fully
contained within the central few spaxels of the pointing.  However, in
case of extended emission spread over more than a single
pointing~($\pp{47} \times \pp{47}$), we used the HIPE inbuilt task
`specProject' for obtaining the integrated line flux. This algorithm
involves a projection of all the positions in the input cube onto the
sky. It calculates a regular RA/Dec grid with a default spaxel size of
$\pp{3}$ and projects the raster positions onto it. The details of
this algorithm can be found in the spectroscopy pipeline described in
the PDRG documentation. Thereafter, we conducted `aperture photometry'
by placing a rectangular box around the visible emission and
integrating the flux to obtain a spectrum. Since none of our
observations were performed in dithered mode, the main necessity of
using `specProject' was to obtain a projection of the multiple single
pointings on the sky plane. Since the individual pointings overlap by
a considerable amount, simply adding the fluxes in all spaxels
contributing above a certain threshold results in an overestimation of
the line emission.

We used the Levenberg-Marquardt minimization routine
\citep{Levenberg1944,Marquardt1963} to fit a model to an observed
spectrum. The model comprised a gaussian for the line emission and a
polynomial of order 0 (or 1) for the
pseudo-continuum\footnote{Presently, there are offsets in the
  continuum levels between the different spaxels preventing the
  extraction of reliable continuum data.}  baseline. A positive line
detection was based in terms of the signal-to-noise ratio~(SNR). The
SNR was defined as the ratio of the line peak to the standard
deviation of the data about the fitted model.

\subsection{PACS Photometry}
\label{pacsphoto}

\begin{figure*}
  \begin{minipage}{0.495\textwidth}
    \centering
    \includegraphics[width=0.8\textwidth]{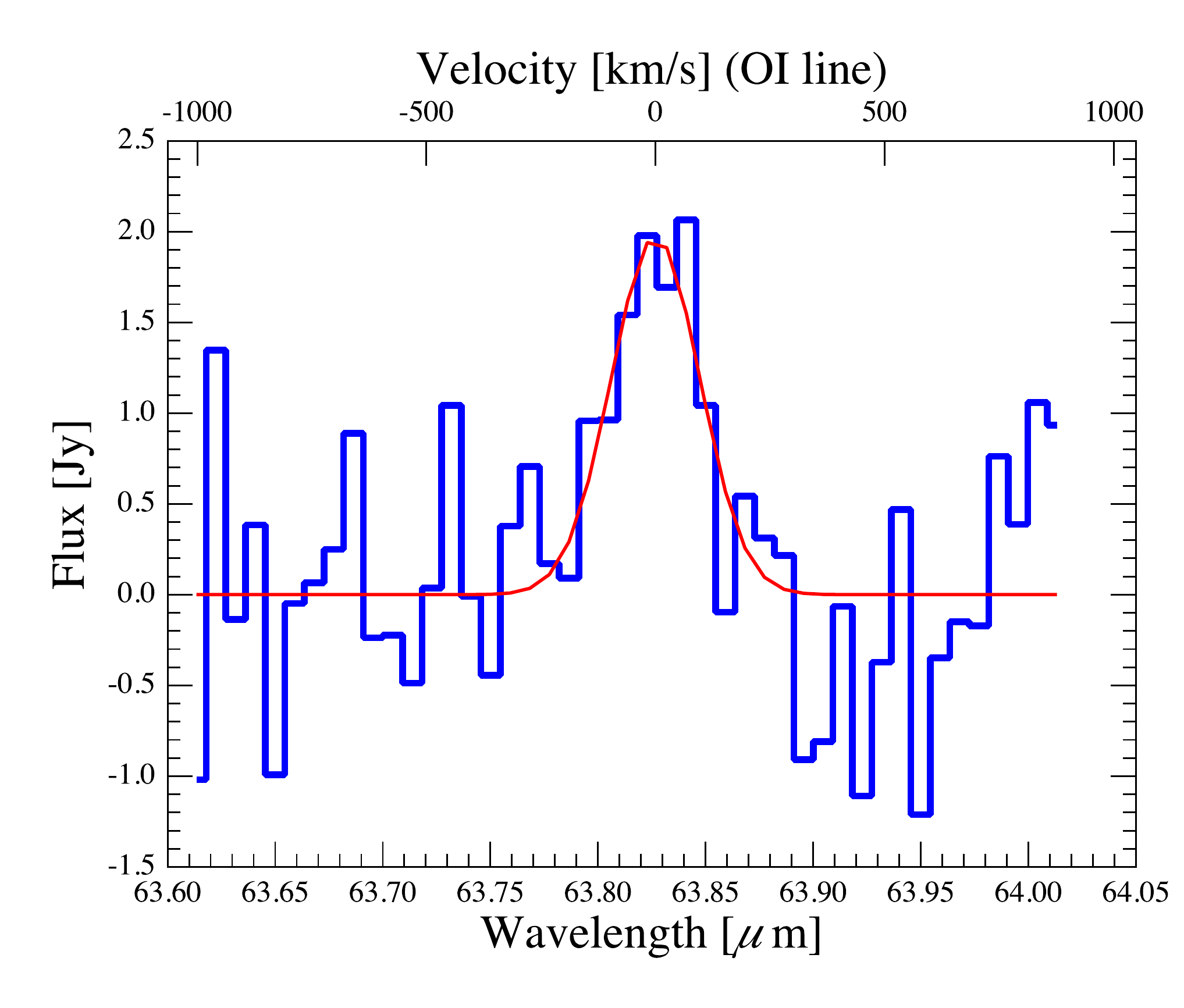}
  \end{minipage}%
  \begin{minipage}{0.495\textwidth}
    \centering
    \includegraphics[width=0.8\textwidth]{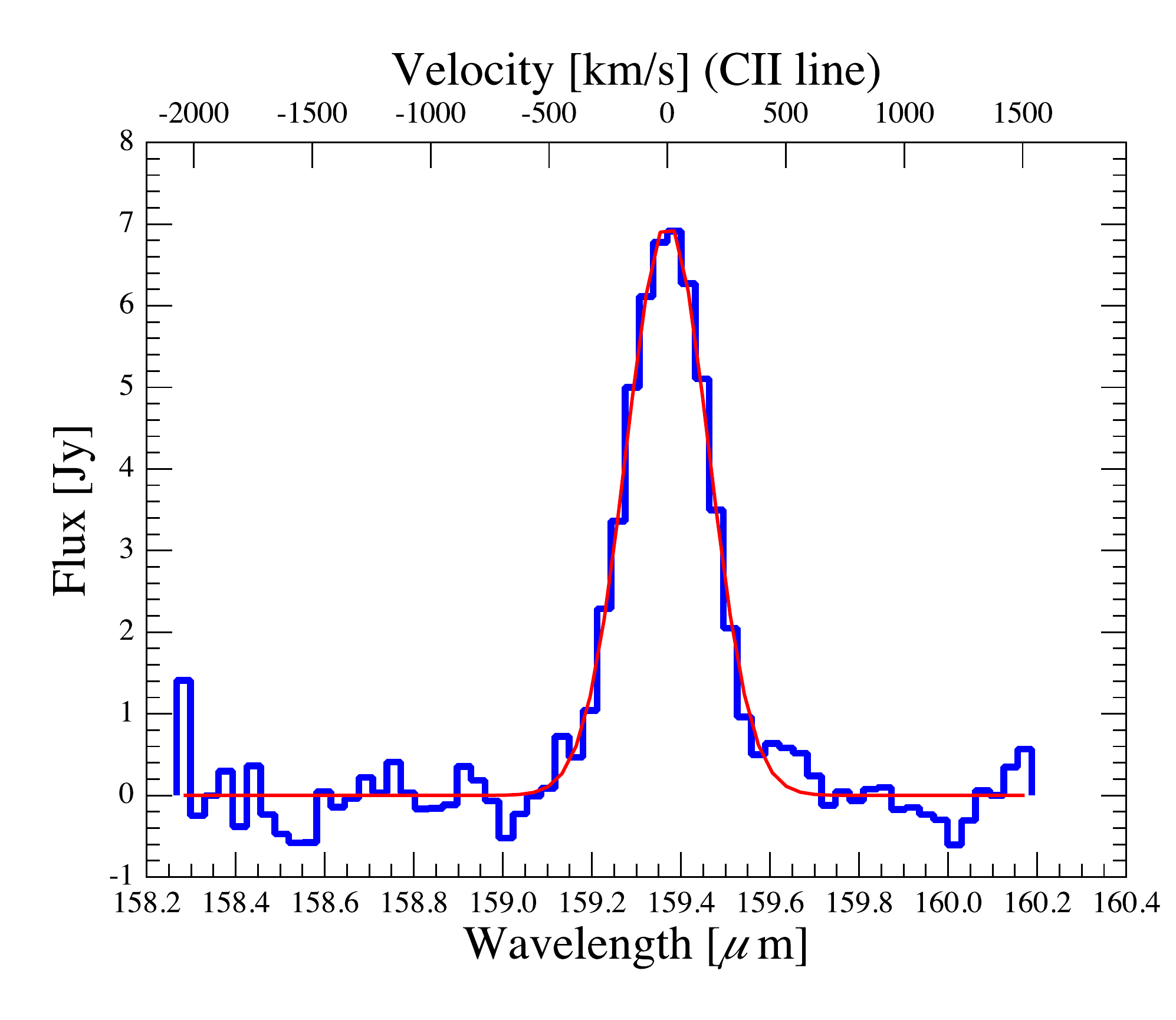}
  \end{minipage}\\
  \begin{minipage}{\textwidth}
    \centering
    \includegraphics[width=0.4\textwidth]{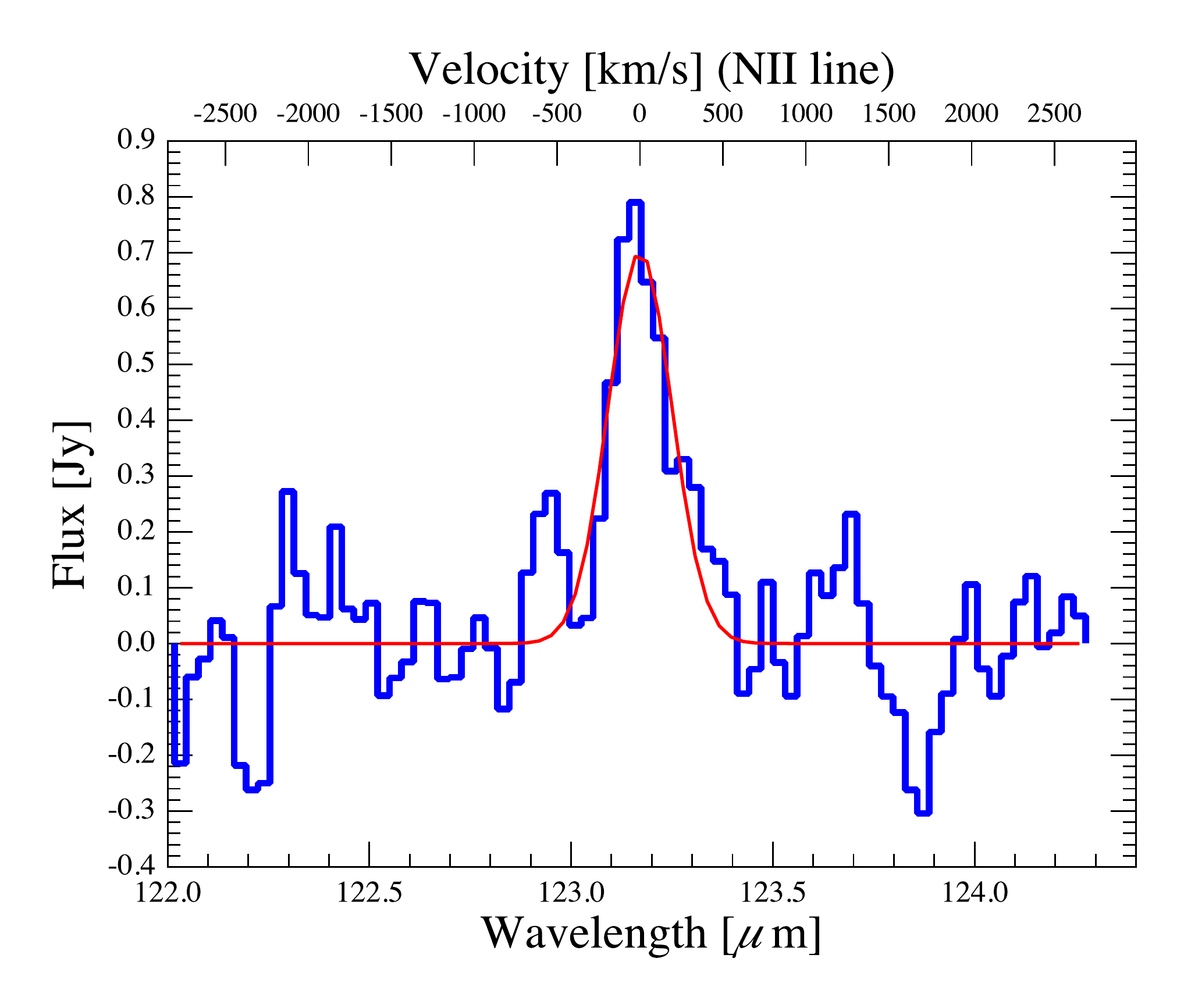}
  \end{minipage}
  \caption{\small The forbidden far-infrared line detections in the
    center of {\cen} made with the {\herschel} PACS instrument. The
    lines are spatially integrated: the {\oi} 63.18{\mm} line~(upper
    left panel, marginally extended), the {\cii} 157.74{\mm}
    line~(upper right panel, extended) and the {\nii}
    121.9{\mm}~(lower panel, point-like). }
  \label{censpec}
\end{figure*}

The PACS photometric observations were taken in large scan mapping
mode in all three bands of the photometer, blue-short~(BS)~(70{\mm}),
blue-long~(BL)~(100{\mm}) and red~(R)~(160{\mm}) using the medium scan
speed ($\pp{20}$s$^{-1}$).  The scan maps comprised 18 scan line legs
of $\p{4}$ length and cross-scan step of $\pp{15}$. Each observation
had a ``scan'' and an orthogonal ``cross-scan'' direction and we
calibrated the corresponding data separately before combining them
into a single map of $\p{9} \times \p{9}$.  The resulting maps have a
resolution of $\pp{5.2}, \pp{7.7}$ and $\pp{12}$ at 70{\mm}, 100{\mm}
and 160{\mm}.  The PACS photometer performs dual-band imaging such
that the BS and BL bands each have simultaneous observations in the R
band so we have two sets of scans in the R band.

For PACS photometry, we used the HIPE version 3.0~CIB~1475 to reduce
the data and adopted the PDRG to process the raw level-0 data to
calibrated level~2 products.  We employed the official script for PACS
ScanMapping mode with particular attention to the high pass filtering
to remove ``1/$\sqrt{f}$'' noise.  We used the `HighPassFilter' method
to remove the large scale ($\gtrsim \pp{80}$) artefacts.  The target
BCG and other bright sources in the field were masked prior to
applying the filter. The size of the mask was chosen to be less than
the filter size so as to minimize any left-over low-frequency
artefacts under the masks. We used a filter size of 20 readouts for
the BS and BL bands and 30 readouts for the R band and a mask radius
of $\pp{20}$ for the BS and BL bands and $\pp{30}$ for the R
band. Finally the task `photProject', was used to project the
calibrated data onto a map on the sky in units of Jy~pixel$^{-1}$. The
``scan'' and ``cross-scan'' maps were then averaged together to
produce the final coadded map.

\begin{table*}
  \centering
  \caption{\small Estimated parameters for the far-infrared forbidden lines in {\cen}. Also given are the upper
    limits on the line flux for the non-detections. The total extent of the {\cii} and {\oi} emission was estimated
    based on visual inspection. }
  \label{cenparam}
  \begin{tabular}{c c c c c c c c c}
    \hline
    Line &  $\lambda$ ($\mu$m)  & \multicolumn{2}{c}{Offset (km~s$^{-1}$)}    &  \multicolumn{2}{c}{FWHM (km~s$^{-1}$)} & Line Flux & Total Extent \\
    &                                    & z$_{\st {bcg}}$ & z$_{\st{cl}}$                  &  Obs.    &  Intrinsic      & ($10^{-18}$) W/m$^2$ &  \\   
    \hline\hline
    {\oi}   &  63.827$\pm $0.005   \comment{& 0.0102}   &  -15$\pm$28  & -52$\pm$28  & 228$\pm$54 & 218$\pm$35 &  57.6$\pm$7.7 & $\pp{15}$  \\ 
    {\cii}  &  159.370$\pm$0.003  \comment{& 0.0103}   &    18$\pm$11  & -22$\pm$11  & 410$\pm$13 & 335$\pm$16 & 174.7$\pm$3.1 & $\pp{35}$  \\
    {\nii}  & 123.171$\pm$0.009   \comment{& 0.0104}   &   48$\pm$40  &    8$\pm$40  & 452$\pm$54 & 351$\pm$69 &  24.9$\pm$1.7  & $<\pp{9.4}$  \\          
    {\oiii}   & ...  & ...  & ...  & ... & ... &  $< 3$ & ... \\          
    {\oib}   & ...  & ...  & ...  & ... & ... &  $< 2$ & ... \\
    {\si}     & ...  & ...  & ...  & ... & ... &  $< 4$ & ... \\                    
    \hline
  \end{tabular}
\end{table*}

\subsection{SPIRE Photometry}
\label{spirephoto}

The SPIRE photometry was performed in the large scan map mode with
cross-linked scans in two orthogonal scan directions.  The photometer
has a field of view of $\p{4} \times \p{8}$, which is observed
simultaneously in three spectral bands, PSW~(250{\mm}), PMW~(350{\mm})
and PLW (500{\mm}) with a resolution of about $\pp{18}$, $\pp{25}$ and
$\pp{36}$, respectively.

For SPIRE photometry, we used a newer HIPE version 4.0~CIB~1432 and
the standard HIPE pipeline for the LargeScanMap observing mode and the
na\"ive map-maker.  The pre-processed raw telemetry data were first
subject to engineering conversion wherein the raw timeline data were
converted to meaningful units, the SPIRE pointing product was created,
deglitching and temperature drift correction were performed, and maps
were created, the units of which were Jy beam$^{-1}$.

\section{{Results}}
\label{results}

\subsection{{Line Detections}}
\label{lines}

Of the lines observed, we detected {\oi}, {\cii} and {\nii} in {\cen}
(Fig.~\ref{censpec}). We did not detect {\si}, {\oiii} and {\oib}
emission lines at a 3$\sigma$ level of $3.0\times10^{-18}$~W/m$^2$ for
{\si} and {\oiii}, and $1.2\times10^{-18}$~W/m$^2$ for {\oib}. The
line parameters are summarized in table~\ref{cenparam}.

The {\nii} emission line at 121.9{\mm} has been detected only in the
central $\pp{9.4}$ spaxel. The {\oi} emission line spectrum at
63.18{\mm}, although noisy, has clearly been detected. Shown in the
top left panel of Fig.~\ref{censpec} is the continuum-subtracted
spectrum integrated over a central region of $\sim \pp{15}$ in
diameter~(3~kpc). Hence, the {\oi} emission is slightly extended. The
spectrum is obtained with oversample=2 and upsample=1 followed by the
projection technique, `specProject', described in
Sect.~\ref{specanalysis2}.

\begin{figure*}
  \centering

  \begin{minipage}{0.33\textwidth}
    \centering
    \includegraphics[width=\textwidth]{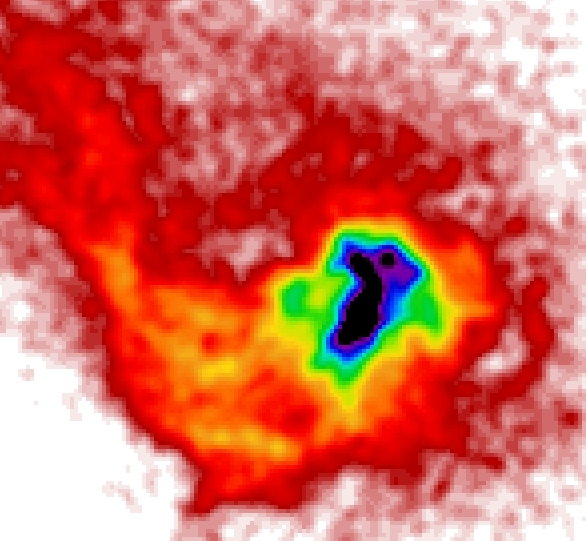}
  \end{minipage}%
  \begin{minipage}{0.33\textwidth}
    \centering
    \includegraphics[width=\textwidth]{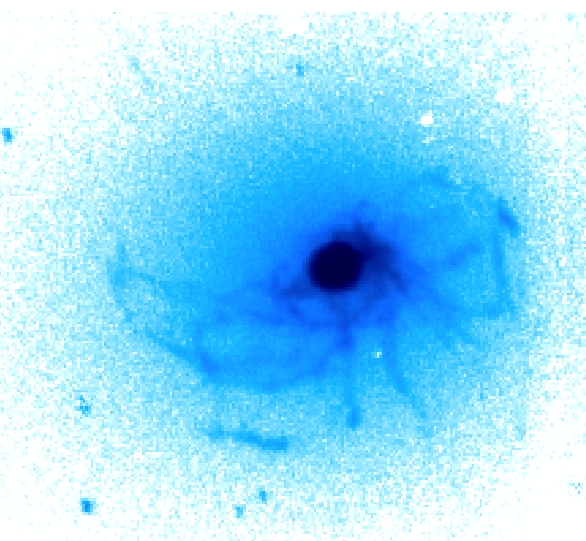}
  \end{minipage}%
  \begin{minipage}{0.33\textwidth}
    \centering
    \includegraphics[width=\textwidth]{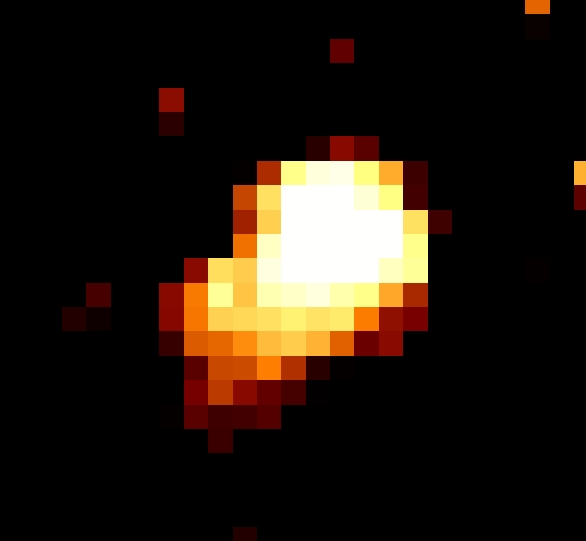}
  \end{minipage}\\[8pt]
  \begin{minipage}{0.5\textwidth}
    \centering
    \includegraphics[width=0.65\textwidth]{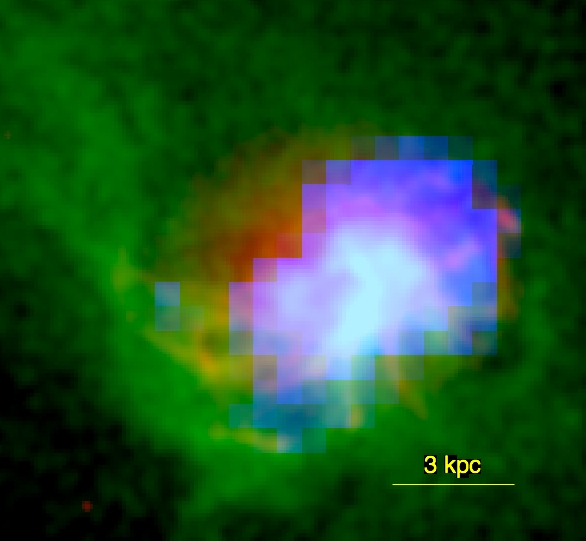}
  \end{minipage}%
  \begin{minipage}{0.5\textwidth}
    \centering
  \includegraphics[width=0.65\textwidth]{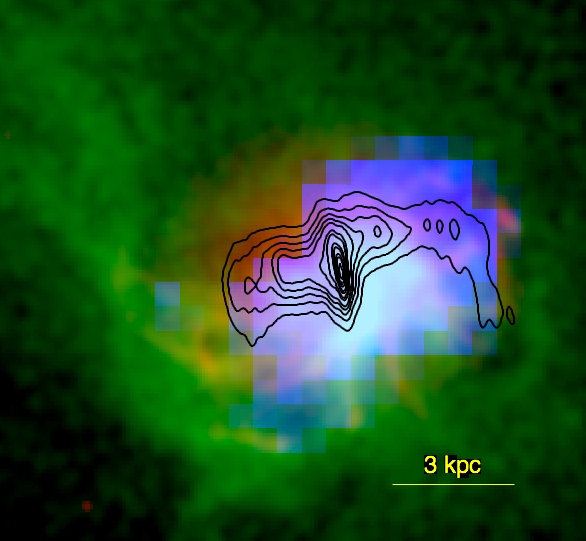}
  \end{minipage}
  \caption{\small {\it Upper panel:} The Chandra X-ray
    surface-brightness map in the soft energy band (0.5 $-$ 1)~keV
    (left), the Gemini optical {\ha} emission (middle)
    \protect\cite{Crawford2005} and the Herschel far-infrared {\cii}
    157.74{\mm} emission (right). {\it: Lower panel} A multiwavelength
    RGB image of {\cen} with contours. {\it Red}: optical {\ha} line
    emission. {\it Green}: X-ray emission, {\it Blue}: far-infrared
    {\cii} emission.  {\it Black contours}: VLA 1.4~GHz radio
    contours~(NRAO/VLA image archive). Each side measures an arcminute
    corresponding to a linear scale of 12~kpc.}
  \label{cencii}
\end{figure*}

The {\cii} continuum-subtracted integrated line emission at
157.74{\mm} is shown in the top right panel of Fig.~\ref{censpec}. It
is obtained using oversample=1 and upsample=4 and projecting the two
nods onto the sky. Though oversample=1 results in poorer instrumental
resolution, it helps in beating down the noise and upsample=2 or 4
helps recover some of the lost information.  The two-dimensional
distribution of {\cii} emission is shown in the multiwavelength
collage in Fig.~\ref{cencii}, which contains an optical {\ha} image
from NTT, an X-ray image from $\chandra$ and 1.4~GHz radio contours
from VLA/NRAO\footnote{The National Radio Astronomy Observatory is a
  facility of the National Science Foundation operated under
  cooperative agreement by Associated Universities, Inc.}. The
\textit{Chandra} datasets used to create the upper left panel of
Fig.~\ref{censpec} were ObsIds 504, 4954, 4955 and 5310, which were
first presented in \cite{Sanders2002} and
\cite{Fabian2002,Fabian2005}.  The archival datasets were reprocessed
with \textsc{ciao}~4.3 and \textsc{caldb}~4.4.0 to apply the
appropriate gain and charge transfer inefficiency correction, then
filtered to remove photons detected with bad grades and cleaned of
flare periods. The final cleaned exposure time was 201.6~ks.  Each of
the final cleaned events files was then reprojected to match the
position of the 4954 observation.  A combined image was produced by
summing images in the (0.5 to 1.0)~keV band extracted from the
individual reprojected datasets.  This image was then corrected for
exposure variation by dividing the summed image by summed exposure
maps created for each dataset.

\begin{figure*}
  \begin{minipage}{0.33\textwidth}
    \centering
    \includegraphics[width=1\textwidth]{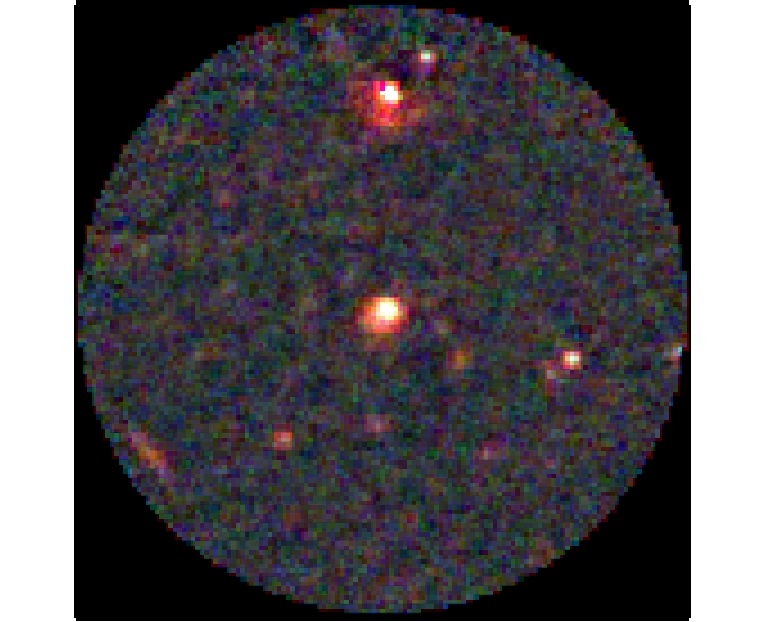}
  \end{minipage}%
  \begin{minipage}{0.33\textwidth}
    \centering
    \includegraphics[width=1\textwidth]{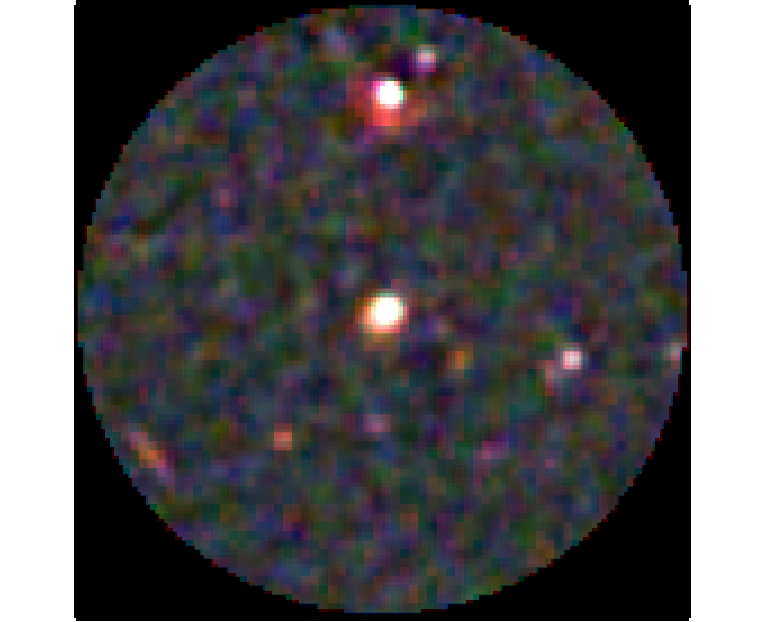}
  \end{minipage}%
  \begin{minipage}{0.33\textwidth}
    \centering
    \hspace*{-0.35cm}
    \includegraphics[width=1.15\textwidth]{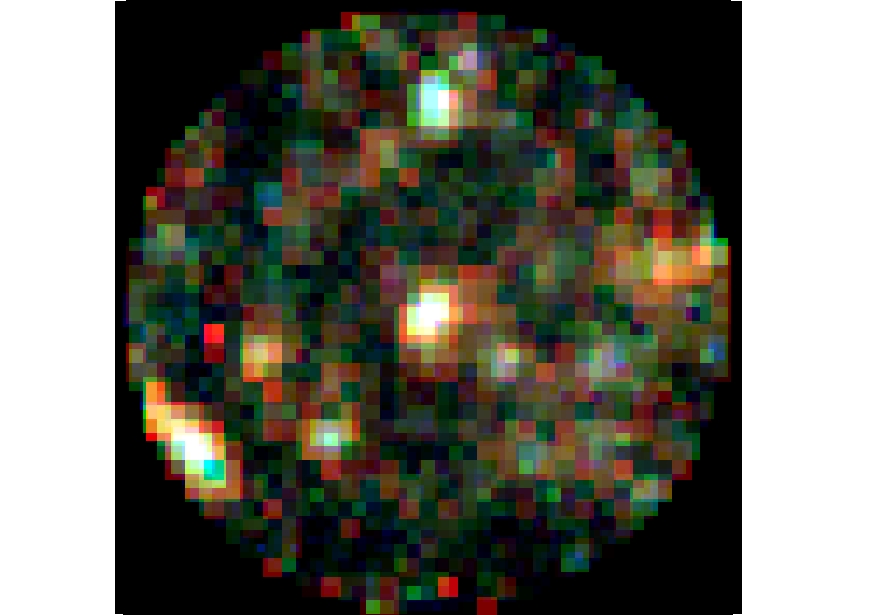}
  \end{minipage}%
  \caption{\small Photometry images of {\cen}. {\it Left}: PACS colour
    images at 70{\mm}~(blue), 100{\mm}~(green) and 160{\mm}~(red) with
    a resolution of $\pp{6}$, $\pp{7}$ and $\pp{12}$
    respectively. {\it Center}: PACS colour images combined with the
    same smoothing gaussian of FWHM $\pp{12}$. {\it Right}: SPIRE
    colour images at 250{\mm}~(blue), 350{\mm}~(green) and
    500{\mm}~(red) with a resolution of $\pp{18}$, $\pp{24}$ and
    $\pp{38}$ respectively. The image units are Jy/beam and each side
    measures $\p{7}$. }
  \label{hpp}
\end{figure*}

The total extent of the {\cii} emission is about $\pp{35}$~(7~kpc) in
diameter. The {\cii} image exhibits a northwest-southeast elongation,
which in conjunction with the optical and X-ray emission maps of the
spiral filaments, is clearly the overall elongation direction
associated with {\cen}. The PACS continuum dust emission shows similar
asymmetry about the center of the galaxy (Fig.~\ref{hpp}). The bulk of
the {\cii} emission appears to be displaced to the west relative to
the {\ha} core. The centroid of the {\cii} emission is indeed offset
to the west of the {\ha} core by about $\pp{5}$ (half spaxel
size). The reported $1\sigma$ absolute pointing accuracy of
{\herschel} is $\pp{2}$ for pointed observations and larger for scan
map observations. As a check, we analysed the {\cii} line observations
with the latest official version of HIPE (6.0.2055) with updated
calibration files and found a similar distribution of the {\cii}
emission with a similar offset from the {\ha} core. While the offset
between the {\cii} and {\ha} emission may also be real and reflect
inhomogeneities in the ISM properties or different excitation
mechanisms, given the similarity between {\cii} and {\ha} it is more
likely that the offset is due to the pointing uncertainty.

\subsection{Aperture Photometry}
\label{photoanalysis}

We conducted aperture photometry using a variety of different methods
with the aim of attaining robust flux measurements and reliable
uncertainties. These methods included routines inbuilt within HIPE,
GAIA~(Graphical Astronomy and Image Analysis Tool) etc. The spread in
the flux estimates was within the random noise associated with the
measurements.

For PACS flux densities, small aperture corrections were applied as
outlined in the PACS ScanMap release note. Further flux calibration on
the derived flux densities was performed to account for the known
overestimation introduced by the ground-based flux calibration by
factors of 1.05, 1.09 and 1.29 in the BS, BL and R bands
respectively. The PACS absolute flux accuracy is within 10~\% for BS
and BL, and better than 20~\% for R. The uncertainties are not
believed to be correlated due to the BS and BL bands being taken at
different times and the R band using a different detector. Since we
used a newer HIPE version to obtain SPIRE images, the known flux
calibration offset was accounted for during the data conditioning and
hence no multiplicative calibration factors were required. The SPIRE
absolute flux accuracy is within 15~\% for all three bands. The flux
measurements from {\herschel} and other instruments are given in
table~\ref{hppflux}.

\begin{table}
  {\small
  \centering
  \caption{\small Compilation of infrared flux-densities.}
  \label{hppflux}
  \begin{tabular}{| c | c | c | c |}
    \hline
    $\lambda$ ($\mu$m)  & Instrument & Aperture  &  Flux (mJy) \\
    \hline\hline
    3.6    & IRAC Spitzer$^{\st a}$       &  $\pp{65}$  &  328$\pm$17 \\
    4.5    & IRAC Spitzer$^{\st a}$       &  $\pp{65}$  &  191$\pm$10 \\
    5.8    & IRAC Spitzer$^{\st a}$       &  $\pp{65}$  &  162$\pm$8 \\
    8.0    & IRAC Spitzer$^{\st a}$       &  $\pp{65}$  &  100$\pm$5 \\
    12     & IRAS$^{\st c}$                   &                    &  $< 25$        \\
    24     & MIPS Spitzer$^{\st a}$       &  $\pp{65}$  &  25.6$\pm$3.0 \\
    24     & MIPS Spitzer$^{\st b}$       &  $\pp{30}$  &  24$\pm$4 \\
    25     & IRAS$^{\st c}$                   &                    &  $< 28$        \\
    60     & IRAS$^{\st c}$                   &                    &  100$\pm$23 \\
    70     & MIPS Spitzer$^{\st a}$       &  $\pp{35}$  &  165$\pm$33 \\
    70     & MIPS Spitzer$^{\st b}$       &  $\pp{30}$  &  156$\pm$19 \\
    70     & PACS Herschel$^{\st d}$   &  $\pp{36}$  &  114$\pm$11 \\  
    100   & PACS Herschel$^{\st d}$   &  $\pp{36}$  &  244$\pm$24 \comment{8} \\ 
    100   & IRAS$^{\st c}$                   &                    &  830$\pm$148 \\
    160   & PACS Herschel$^{\st d}$   &  $\pp{36}$  &  383$\pm$77  \comment{9} \\ 
    160   & MIPS Spitzer$^{\st a}$       &  $\pp{40}$  &  256$\pm$51 \\
    160     & MIPS Spitzer$^{\st b}$       &  $\pp{30}$  &  331$\pm$29 \\
    250   & SPIRE Herschel$^{\st d}$   &  $\pp{38}$  &  208$\pm$30 \\ 
    350   & SPIRE Herschel$^{\st d}$   &  $\pp{38}$  &  98$\pm$15 \\ 
    500   & SPIRE Herschel$^{\st d}$   &  $\pp{38}$  &  38$\pm$6 \\ 
    \hline 
    \multicolumn{4}{p{8cm}}{ $^{\st a}$ The {\spitzer} IRAC/MIPS data are from a sub-sample of the 
      ACCEPT sample of galaxy clusters \citep[][ in preparation]{Hoffer2011}. The errorbars reflect
      the absolute flux uncertainties.} \\
    \multicolumn{4}{p{8cm}}{$^{\st b}$ \cite{Kaneda2005,Kaneda2007}.}\\
    \multicolumn{4}{p{8cm}}{$^{\st c}$ The {\iras} data are from NASA/IPAC 
      Extragalactic Database (http://nedwww.ipac.caltech.edu).}\\
    \multicolumn{4}{p{8cm}}{$^{\st d}$ Herschel PACS/SPIRE data are from this work. The 
      Herschel PACS/SPIRE errorbars reflect the absolute flux uncertainties: 10\% at PACS BS and BL
      and 20\% at R, and 15\% at all SPIRE wavelengths.}\\
  \end{tabular}
}
\end{table}

Our Herschel flux estimate disagrees with the Spitzer MIPS 160{\mm}
measurement of \cite{Hoffer2011}, though it is consistent with the
measurement of \cite{Kaneda2005}.  A systematic error in the
performance of either of the two instruments can be ruled out since
while at 70{\mm} the Spitzer flux estimate is higher than the Herschel
estimate, at 160{\mm} the situation is reversed. In addition, we also
ruled out the discrepancy being due to the difference in the spectral
response functions of Herschel PACS and Spitzer MIPS photometers. We
used the above fitted SED of {\cen} as the `true' source model and
calculated the predicted flux density for each instrument as the
weighted average, where the weights were determined from the spectral
response function of each instrument. The Herschel PACS and Spitzer
MIPS predicted flux densities at 70{\mm} and 160{\mm} estimated this
way are consistent with each other to better than 5~\%. Since the
Spitzer MIPS 160{\mm} flux shows a spread between the measurements
from \cite{Kaneda2005} and \cite{Hoffer2011}, we give preference to
the Herschel estimates, especially since {\it Herschel} is designed to
operate at higher sensitivity at FIR wavelengths.

\begin{figure*}
  \begin{minipage}{0.495\textwidth}
    \centering
    \includegraphics[width=0.8\textwidth]{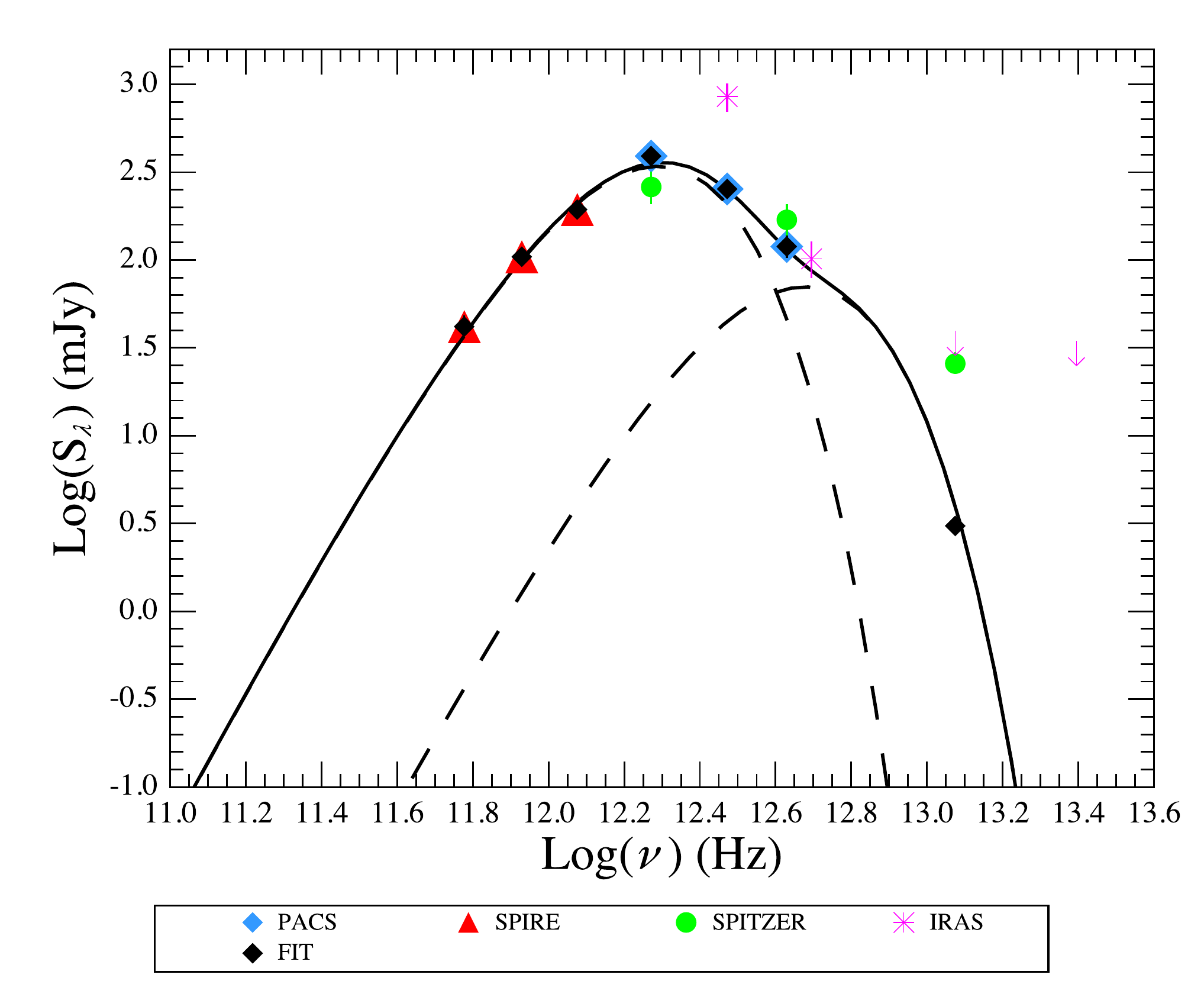}
  \end{minipage}%
  \begin{minipage}{0.495\textwidth}
    \centering
    \includegraphics[width=0.8\textwidth]{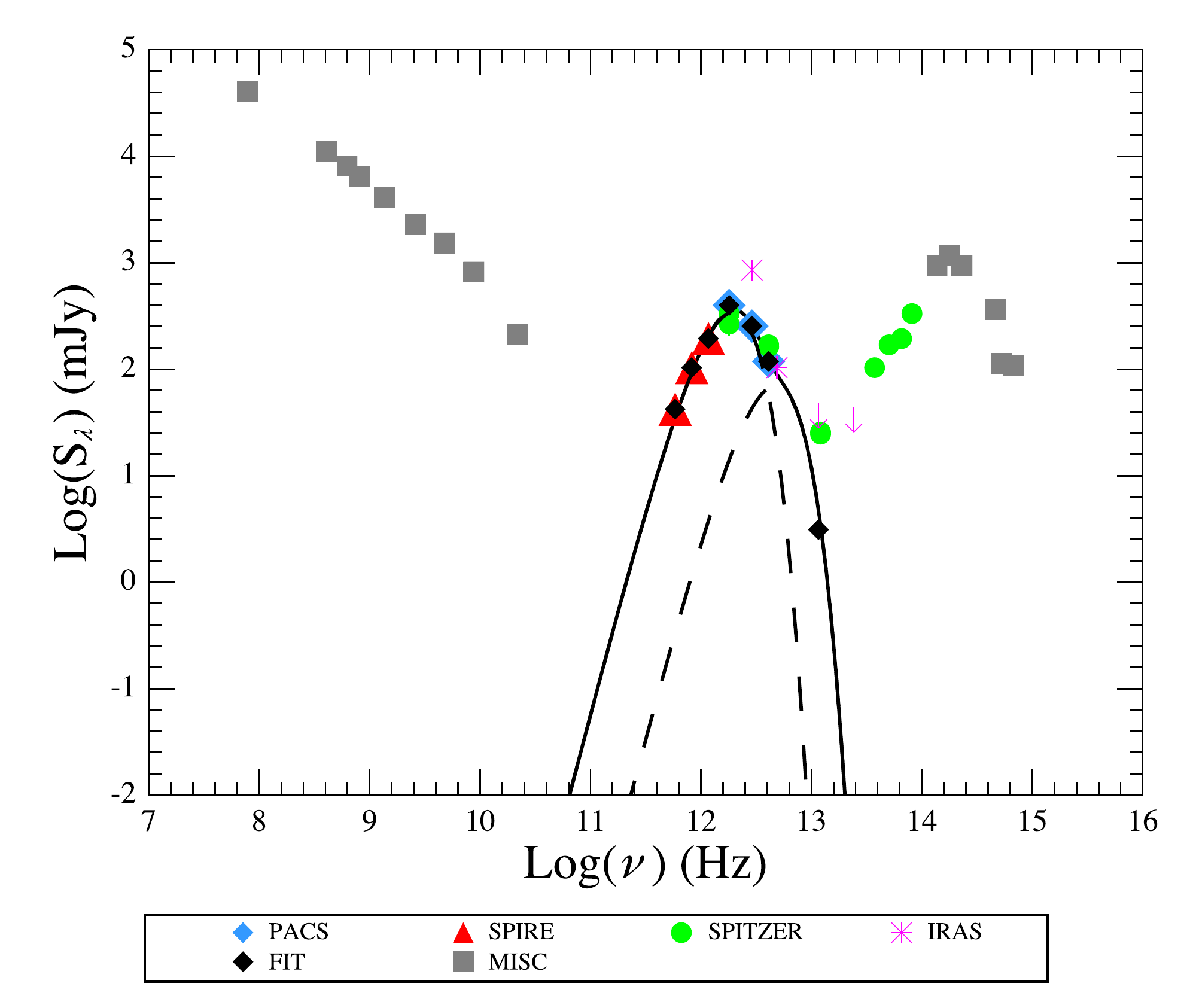}
  \end{minipage}
  \caption{\small The spectral energy distribution~(SED) of
    {\cen}. {\it Left}: The far-infrared SED obtained from a modified
    two-component blackbody fit to the Herschel and Spitzer data,
    designated as `FIT' (filled black diamonds). {\it Right}: The
    complete SED extending to radio frequencies on one end and optical
    and X-ray frequencies on the other, collectively designated as
    `MISC'.}
  \label{sed}
\end{figure*}

\subsection{Dust SED}
\label{dust}

The dust emission from {\cen} has been detected in all six PACS and
SPIRE bands. In Fig.~\ref{hpp}, we show the PACS images at 70{\mm},
100{\mm} and 160{\mm} and the SPIRE images at 250{\mm}, 350{\mm} and
500{\mm} centered on {\cen}. The PACS photometry data reduction
pipeline described in Sect.~\ref{pacsphoto} produces images with units
Jy~pixel$^{-1}$. We converted the units from Jy~pixel$^{-1}$ to
Jy~beam$^{-1}$ by using the conversion factor $\pi/(4\ln2)
(\theta_{\st{beam}}/\theta_{\st{pix}})^2$~pixel~beam$^{-1}$, where
$\theta_{\st{beam}}$ is $\pp{6}$, $\pp{7}$ and $\pp{12}$ and
$\theta_{\st{pix}}$ is $\pp{3.2}$, $\pp{3.2}$ and $\pp{6.4}$ for the
70{\mm}, 100{\mm} and 160{\mm} bands, respectively.

The BCG emission seems to be slightly extended along the
northwest-southeast direction. The origin of the infrared continuum
emission is the interstellar dust grains in the host galaxy, which are
heated by the absorption of photons from an ionizing source. Whether
this ionizing source is starbursts~(SB) or an AGN embedded in the host
galaxy is usually a matter of debate but the AGN/SB diagnostics for
{\cen} suggest the AGN contribution to be very low
(Sect.~\ref{agn}). We fitted the dust spectral energy
distribution~(SED) with a blackbody spectrum modified by a
frequency-dependent emissivity. Following \cite{Aravena2008}, and
references therein, the flux-density at a given frequency, $\nu$, is :
\begin{equation}
  S_{\nu} = \frac{\Omega}{(1+z)^3} \, \left[ B_{\nu}(\td) - B_{\nu}(\tcmb) \right] \, (1- e^{-\tau_{\nu}(\md)}),
\end{equation}
where $B_{\nu}(T)$ is the Planck function at frequency $\nu$, and
temperature $T$. $B_{\nu}(\tcmb)$ is the contribution from the cosmic
microwave background at $\tcmb = 2.73(1+z)$~K. $\Omega$ is the solid
angle subtended by the source, $\pi \theta^2$, where $\theta$ is the
source size in radians (here assumed to be the resolution of the PACS
photometer at 70{\mm}). The last term in parenthesis is the
modification to the blackbody radiation spectrum, where $\tau_{\nu}$
is the dust optical depth,
\begin{equation}
\tau_{\nu} = \kappa_{\nu} \frac{\md}{\Da^2 \Omega}.
\end{equation}
Here, $\kappa_{\nu} =
5.6\times(\nu/3000~\st{GHz})^{\beta}$~m$^2$~kg$^{-1}$
\citep{Dunne2000} is the dust absorption coefficient and $\beta$ is
the dust emissivity index which we fixed to 2 .

Motivated by the findings of \cite{Dunne2001}, we fitted the infrared
data with two modified blackbody functions, representing a cold dust
component and a warm dust component.  This gives four model
parameters: the temperature and mass of the cold dust component,
$\tdc$, $\mdc$, and the temperature and mass of the warm dust
component, $\tdw$, $\mdw$.  We used the Levenberg-Marquardt
minimization routine from Numerical Recipes to obtain the best fit
model parameters. We conducted the fit using seven data points: the
six PACS and SPIRE photometer data and the Spitzer MIPS data at
24{\mm}. The Spitzer MIPS flux density at 24{\mm} receives a
contribution from the hot dust component as well as the passive
stellar population. In order to use the 24{\mm} data for the SED
fitting, it is necessary to subtract the contribution from the passive
stellar population. For this we fitted a powerlaw to the IRAC data,
since the IRAC wavelengths are sensitive to emission from the old
stellar population, and extrapolated the fit to 24{\mm}. The net
flux-density after subtracting the contribution from the passive
stellar population is $3.6\pm3.0$~mJy, which is significantly lower
than the total observed value. It is this net flux-density that we
assigned to the hot dust component for the SED fitting.

The SED fitting yields a massive cold dust component with $\tdc =
(18.9\pm0.7)$~K and $\mdc=(1.6\pm0.3)\times10^6~\ms$ and a low-mass
warmer dust component with $\tdw = (46.0\pm5.0)$~K and
$\mdw=(4.0\pm2.8)\times10^3~\ms$. As also seen in other studies, the
cold component dominates the mass budget. The far-infrared and the
total SEDs, the latter extending from radio to X-ray frequencies, are
shown in Fig.~\ref{sed}.  We used the fit parameters to estimate the
total far-infrared luminosity of the dust between 8{\mm} and
1000{\mm}, $\lfirtot$, and found it to be (7.5$\pm$1.6)$\times
10^8~\ls$.  For the analysis of the photodissociation regions
described in Sect.~\ref{pdr}, we used the FIR luminosity in the range
40{\mm} to 500{\mm}, which from here on is denoted by $\lfir$ and is
equal to (6.2$\pm$0.8)$\times 10^8~\ls$.

To assess the robustness of the derived parameters and the quantities
derived from them, such as the FIR luminosity, we also fitted the
above model using the average of the two MIPS and one PACS flux
estimates at each 70{\mm} and 160{\mm}.  The fit results lead to a
cold dust component with $\tdc = (18.2\pm0.8)$~K and
$\mdc=(1.8\pm0.3)\times10^6~\ms$ and a warm dust component with $\tdw
= (43.0\pm4.7)$~K and $\mdw=(9.1\pm6.8)\times10^3~\ms$. The inferred
FIR luminosities are $\lfirtot=(7.9\pm2.5)\times 10^8~\ls$ and
$\lfir=(6.6\pm1.1)\times 10^8~\ls$. Hence, a $\sim$20\% variation in
the PACS flux densities results in about a 5\% to 10\% variation in
the total cold dust mass and the FIR luminosities. This is well within
the formal errorbars obtained from either of the two fitting methods.

\subsection{Gas-to-Dust Mass Ratio}
\label{gdmr}

\cite{ODea1994} searched for CO in a sample of five radio galaxies,
including {\cen}, using the Swedish-ESO submillimeter telescope.  They
reported nondetection in {\cen} and used the result to put limits on
the gas mass. From these observations, the 3$\sigma$ upper limit on
the mass in the molecular gas is estimated to be $\sim
5\times10^8$~$\ms$. Using the dust mass calculated in
Sect.~\ref{dust}, the (molecular) gas-to-dust mass ratio has an upper
limit of $\sim 325$. Note that the upper limit on the {\it total}
gas-to-dust mass ratio is likely larger since the total-to-molecular
gas mass ratio is greater than unity and may be as high as 5 (or more)
depending upon parameters such as the ionization radiation intensity
and the total hydrogen column density.

We also used the formalism laid out by \cite{Wolfire1990}, which uses
the observed {\cii} emission to obtain the atomic gas mass, assuming
that the {\cii} emission is optically thin:

\begin{eqnarray}
  \mg & = & 2.7\times10^6 \left(\frac{\Dl}{1~\st{Mpc}}\right)^2 \left( \frac{F_{\textrm{C{\sc ii}}}}{10^{-17}~\st{W~cm^{-2}}}  \right) \\ \nonumber
 &  & \times \left(\frac{10^{-21}~\st{ergs~s^{-1}~sr^{-1}~atom^{-1}}}{\Lambda(\textrm{C{\sc ii}})}\right) \times \left(\frac{3\times10^{-4}}{x(\textrm{C{\sc ii}})}\right) \ms
\label{wolfire}
\end{eqnarray}
where $\Dl$ is the luminosity distance, $F_{\textrm{C{\sc ii}}}$ is
the {\cii} line flux, $\Lambda$(C{\sc ii}) is the cooling rate of
ionized carbon and $x$(C{\sc ii}) is the abundance of ionized carbon
relative to hydrogen. Most of the {\cii} emission originates from the
surface layers of the photodissociation regions with the visual
extinction, $A_{\st V} < 4$, depleting steadily with increasing
$A_{\st V}$, and is the dominant of the three most important
carbon-bearing species, C, C{\sc ii} and CO
\citep{Hollenbach1991}. Hence the C{\sc ii} abundance can be equated
to the elemental abundance of carbon, which, for the simulations of
the photodissociation regions described in Sect.~\ref{pdr}, equals 2.5
times the interstellar value, $x$(C{\sc ii}) = $6.3\times10^{-4}$.
For gas temperatures higher than 100 Kelvin (the excitation
temperature of C{\sc ii} is 92~K) and gas densities between
100~cm$^{-3}$ and 1000~cm$^{-3}$, $\Lambda(\textrm{C{\sc ii}})$ is in
the range (10$^{-21}$ --
10$^{-22}$)~$\st{ergs~s^{-1}~sr^{-1}~atom^{-1}}$. Substituting these
values into Eqn.~\ref{wolfire} gives a gas mass in the range (0.5 to
5)~$\times 10^7 \ms$. This is consistent with the upper limit derived
from the non-detection of CO by \cite{ODea1994}.
  
The gas-to-dust mass ratio based on the above estimate of the gas mass
is between 1 and 70. In comparison with the gas-to-dust mass ratios
derived for other BCGs \citep{Edge2001,Edge2010b}, which have typical
values of a few hundred, the ratio in {\cen} is significantly
lower. Note that the above method of determining the gas mass is not
very robust for low $n$ and $G0$, which is the case of {\cen} (see
Sect.~\ref{pdr}), and the dust mass is very sensitive to the assumed
dust temperature, so the derived range of the gas-to-dust mass ratio
should only be treated as a crude estimate.

\subsection{Star Formation Rates}
\label{sfr}

Assuming the dust heating to be due to the young stellar population
and high dust opacity in the star forming regions, the Kennicutt
relation \citep{Kennicutt1998} can be used to convert the $\lfirtot$
(derived in Sect.~\ref{dust}) into a star formation rate~(SFR). Using
the conversion, $SFR~(\mpy) = 4.5\times10^{-44}
\lfirtot~$(erg~s$^{-1}$), and the derived $\lfirtot$, we obtained a
star formation rate of $\sim 0.13~\mpy$. However, we expect a
non-negligible contribution also from the general stellar radiation
field, including the old stellar population in the host galaxy, which
will lower the estimated SFR. Hence, the above value is an upper
limit.

We also retrieved far ultraviolet~(FUV) data from the HST archive. The
data (proposal ID 11681) were taken with the Advanced Camera of
Surveys Solar Blind Channel (ACS/SBC) providing a field of view of
$\pp{34.6} \times \pp{30.5}$.  These observations were made with two
long-pass filters, F150LP and F165LP, and we chose the F150LP filter
over F165LP to estimate the star formation rate (described below) due
to its higher system throughput.

\begin{figure}
    \centering
    \includegraphics[width=0.55\textwidth]{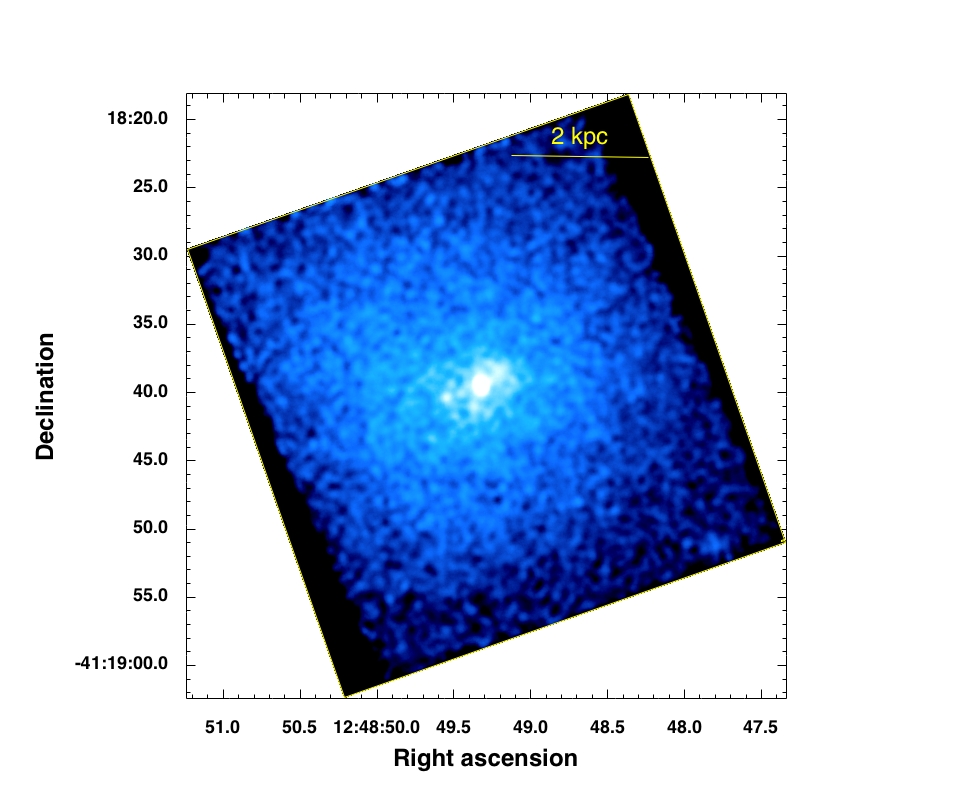}
    \caption{\small The ACS/SBC FUV F150LP observations of {\cen} with
      the {\it HST} showing a core surrounded by some diffuse
      low-brightness emission. The extent of the FUV emission is very
      small relative to {\ha} and {\cii}. Weak FUV emission is
      consistent with the low star formation rate derived in
      Sect.~\ref{sfr} and small values of the FUV intensity field,
      $G0$, described in Sect.~\ref{pdr}.}
  \label{hst}
\end{figure}

We used a Starburst99 stellar library \citep{Leitherer1999} to model a
spectrum of a young stellar population (whose oldest stars are
$10^7$~yr old) with a continuous star formation rate of $0.1~\mpy$. We
fed the output spectrum into the {\it synphot} synthetic photometry
package to determine the expected flux in the F150LP band.  The ratio
of the extinction-corrected observed flux to the expected flux from
synphot multiplied by the assumed SFR in Starburst99 gives a measure
of the actual star formation rate. The FUV emission in the F150LP
band, shown in Fig.~\ref{hst}, contains a bright core surrounded by
low brightness emission. The low surface-brightness extended emission
is most probably dominated by the passive stellar population. Hence,
we derived an upper limit to the SFR based on the assumption that the
entire observed FUV emission arises due to a young stellar population
and a lower limit based on the assumption that only the bright core is
associated with an ongoing star formation. The measured flux using
aperture photometry is $1.23 \times
10^{-15}~$ergs~s$^{-1}$~cm$^{-2}~\AA^{-1}$ if a region of radius $\sim
1.5~$kpc in considered, which encompasses the bright core as well as
the diffuse emission surrounding it, and $2.75 \times
10^{-17}~$ergs~s$^{-1}$~cm$^{-2}~\AA^{-1}$ if only the bright core of
radius $\sim 0.5$~kpc is considered. These values were corrected for
Galactic extinction based on the measured $E(B-V)=0.113$ from NED and
the extinction law, \mbox{$R_V=3.1 [\equiv A(V)/E(B-V)]$}
\citep{Cardelli1989}. The extinction was calculated using the mean
$R_V$ dependent extinction law from \cite{Cardelli1989} in the UV and
FUV regime at the pivot wavelength, \mbox{$\lambda=1612.236~\AA$}. The
extinction corrected fluxes are $2.80 \times
10^{-15}~$ergs~s$^{-1}$~cm$^{-2}~\AA^{-1}$ and $6.27 \times
10^{-17}~$ergs~s$^{-1}$~cm$^{-2}~\AA^{-1}$. Note that there is likely
to be internal extinction within {\cen} and the SFR limits derived
below may increase once this additional extinction is taken into
account.  Comparing the expected flux from the Starburst99 model to
the observed corrected flux gives an upper limit of the FUV derived
star formation rate of $0.08~\mpy$ and a lower limit of $0.002~\mpy$.
The upper value is in agreement with the upper limit derived from the
far-infrared analysis. This is also consistent with the GALEX
near-ultraviolet~(NUV) observations indicative of a SFR consistent
with zero and a 3$\sigma$ upper limit of 0.17~$\mpy$. The GALEX
analysis takes into account the UV-upturn from an evolved stellar
population. The UV excess is calculated by comparing the GALEX NUV
emission to the
2MASS\footnote{http://irsa.ipac.caltech.edu/Missions/2mass.html}
$K$-band emission within an aperture of $\pp{35}$ and attributed to
the ongoing star formation.

The classical X-ray mass deposition rate at a radius of about
$\pp{35}$ from the X-ray peak in the Centaurus galaxy cluster
\citep[the projected separation between the X-ray peak and the BCG is
$\pp{6}$, ][]{Mittal2009} is about (5 to 9)$~\mpy$
\citep[e.g.][]{Sanders2008b,Hudson2010}.  The cooling rates observed
in the form of star formation rate derived from FIR and UV are smaller
by about two orders of magnitude.  Similarly, \cite{Sanders2008b} used
XMM-Newton RGS observations to study the X- ray emission from the core
of the Centaurus galaxy cluster. They carried out detailed spectral
fitting to measure the amount of gas cooling at the center and deduced
an upper limit of $0.8~\mpy$ below 0.4~keV. This is the standard
``cooling-flow'' discrepancy, which is resolved when we consider the
AGN regulated feedback. The luminosity of the gas cooling in the
center can be estimated using a cooling flow model in which the gas
cools from the virial temperature to a minimum possible lower
temperature (0.08~keV).  The cooling luminosity is given by:

\begin{equation}
\lc = \frac{5}{2} \frac{\ms}{\mu m_{\st p}} k\tvir,
\end{equation}

where $\ms$ is the classical mass deposition rate inferred from X-ray
observations, $m_{\st p}$ is the mass of the proton, $\mu$ is the mean
molecular weight ($\sim 0.65$) and $\tvir$ is the cluster virial
temperature which for Centaurus is about 4~keV. Inserting the
estimated $\ms$ and $\tvir$ gives a cooling luminosity in the range
(4.5 to 8.0)~10$^{42}$ erg~s$^{-1}$. On the heating front,
\cite{Rafferty2006} found the mechanical power associated with the
AGN-excavated cavities to be 7.4$\times10^{42}$~erg~s$^{-1}$.
\cite{Merloni2007} estimated a similar value for the mechanical power
$\sim 7.8\times10^{42}$~erg~s$^{-1}$.  Hence there is enough energy in
the radio source to balance cooling of the hot gas and maintain
equilibrium.

\section{{Analysis}}
\label{analysis}

\begin{figure*}
  \begin{minipage}{0.495\textwidth}
    \centering
    \includegraphics[width=\textwidth]{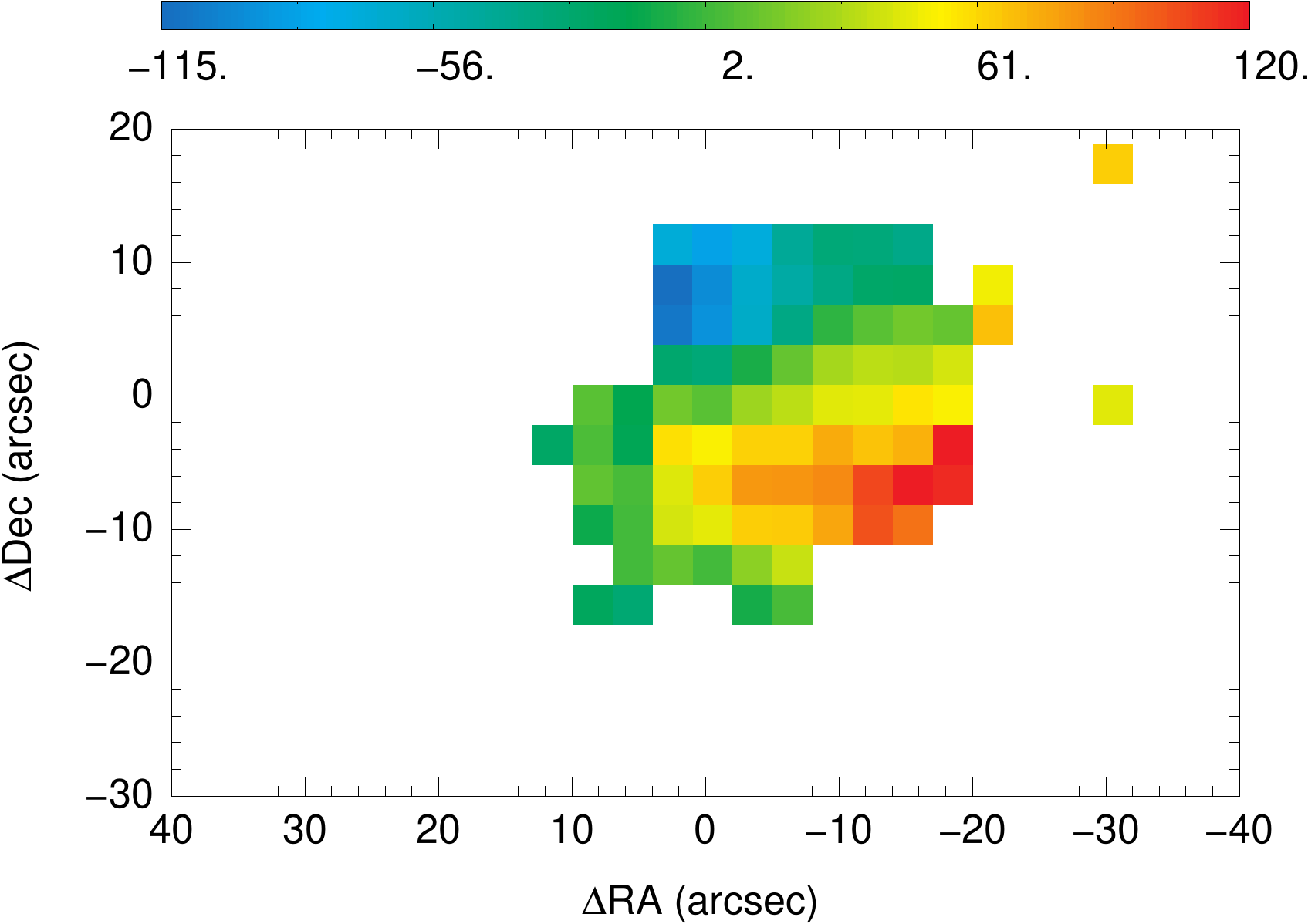}
  \end{minipage}%
  \begin{minipage}{0.495\textwidth}
    \centering
    \includegraphics[width=\textwidth]{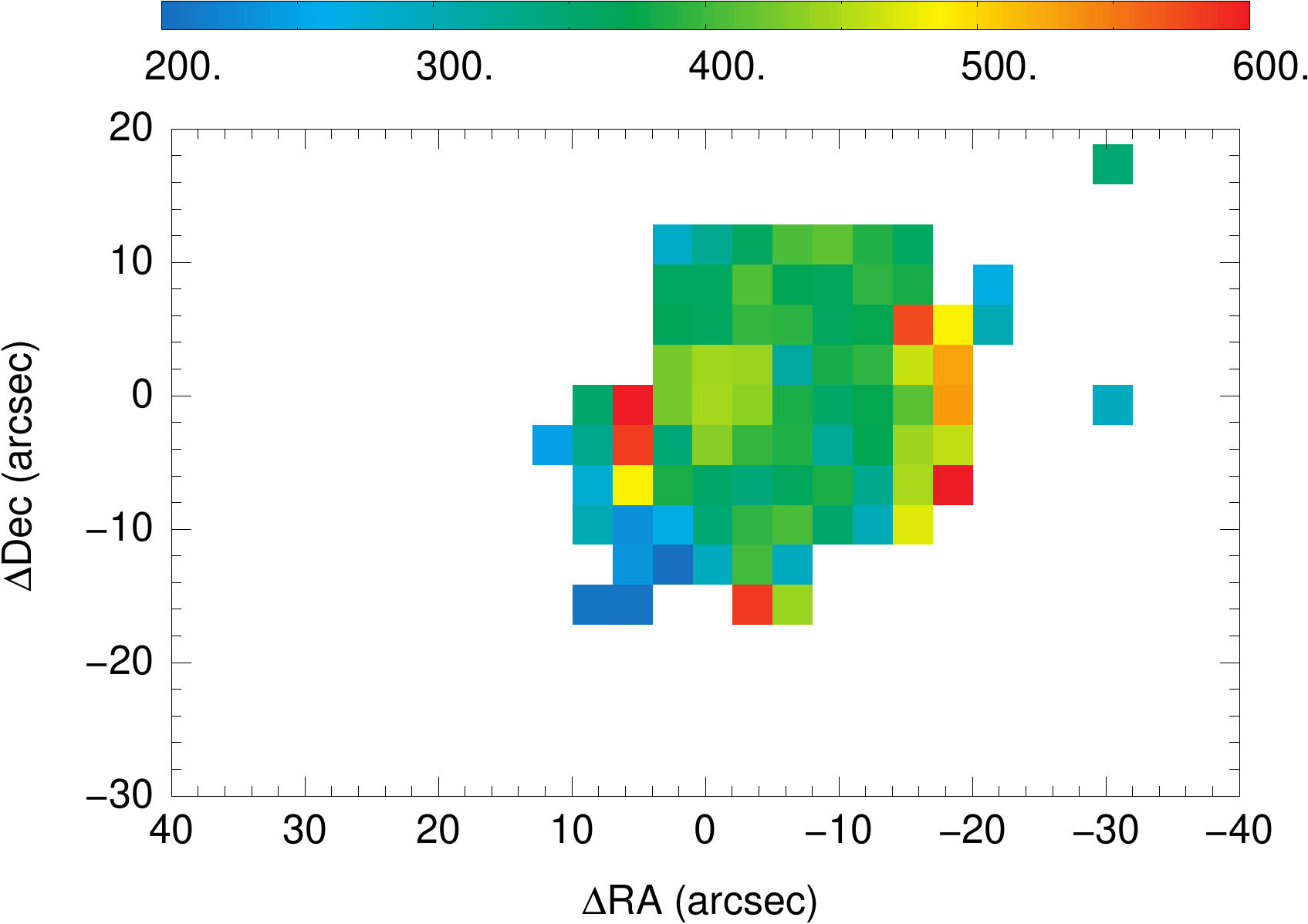}
  \end{minipage}\\
  \begin{minipage}{0.495\textwidth}
    \vspace*{1.5cm}
    \centering
    \includegraphics[width=\textwidth]{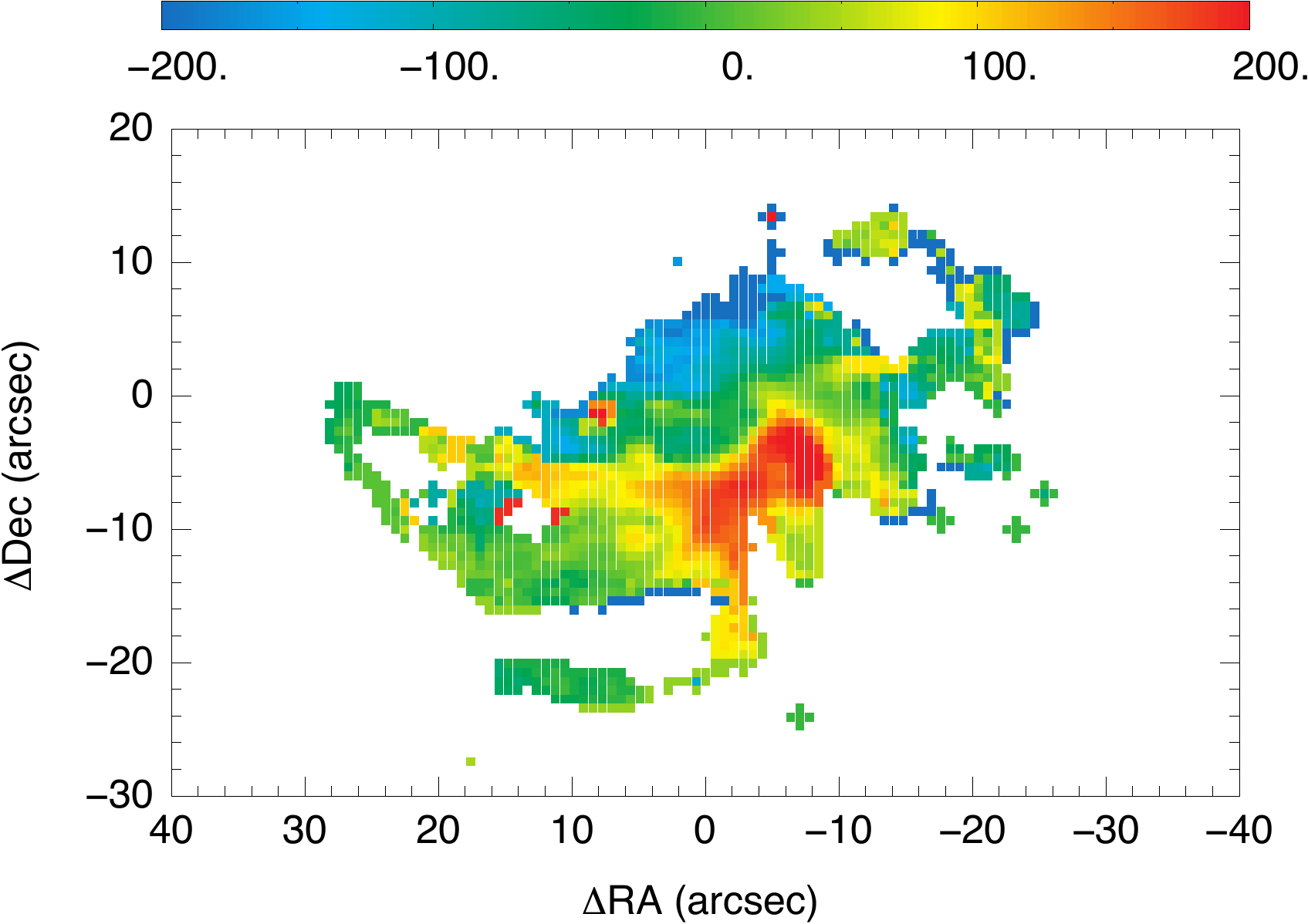}
  \end{minipage}%
  \begin{minipage}{0.495\textwidth}
    \vspace*{1.5cm}
    \centering
    \includegraphics[width=\textwidth]{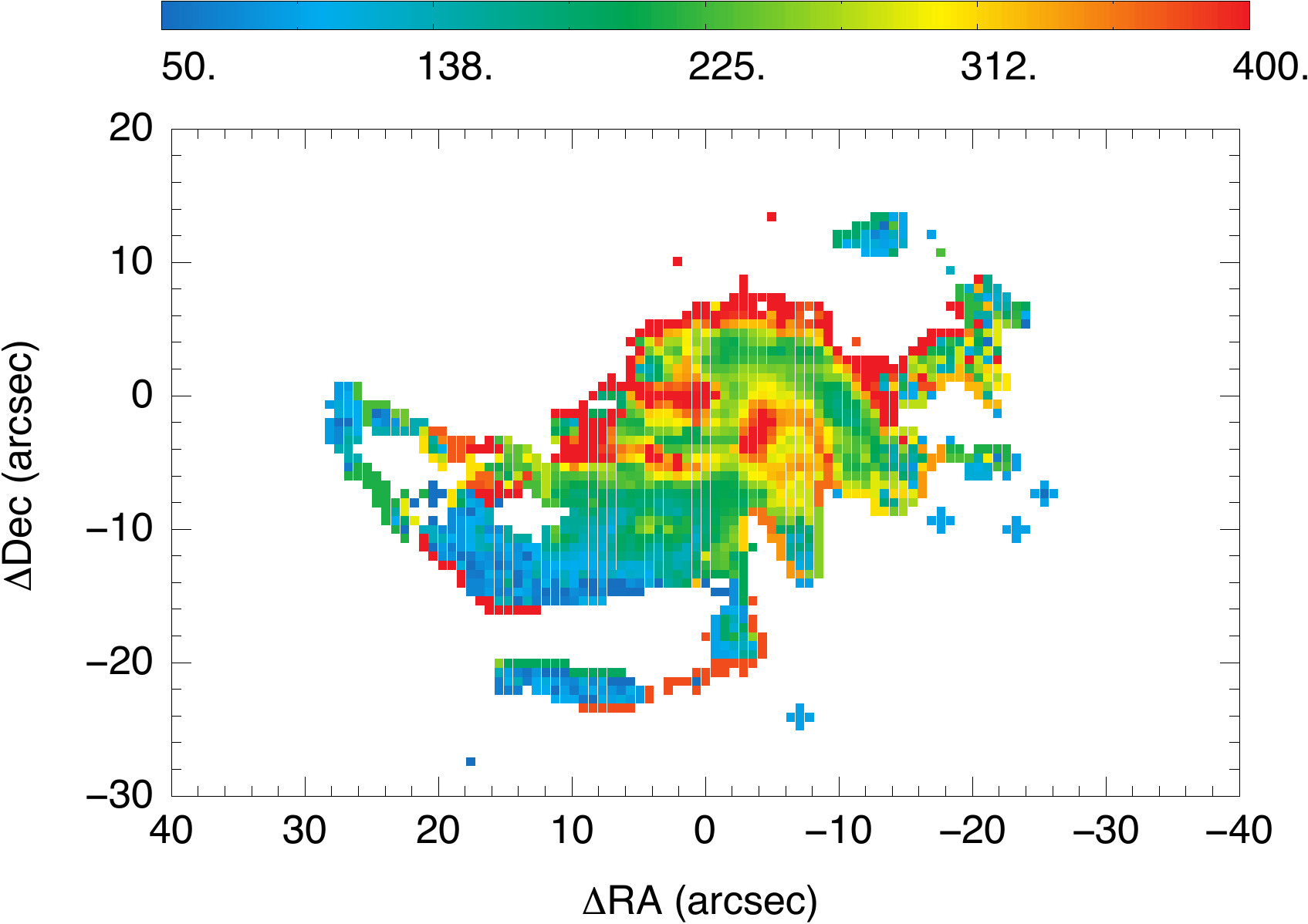}
  \end{minipage}\\
  \caption{\small Gas kinematics. The measured line-of-sight
    velocities (left panels) and line FWHM (right panels) in
    km~s$^{-1}$ of the far-infrared {\cii} gas with a scale
    $\pp{3}$~per~pixel (upper panel) and the optical {\ha}+{\niiopt}
    gas with scale $\pp{0.67}$~per~pixel (lower panel). The
    velocity-dispersion maps have been corrected for the instrumental
    resolution (235~km~s$^{-1}$ for {\cii} and 42~km~s$^{-1}$ for
    {\ha}). The FIR {\cii} emission traces the optical {\ha} emission
    morphologically and kinematically.}
  \label{kinematics}
\end{figure*}

\subsection{AGN contamination}
\label{agn}
 
The interstellar dust is believed to be heated by two principle
mechanisms, by young stars and AGN. Hence it may be that some of the
infrared emission detected is related to the AGN activity.  In order
to gauge the AGN contribution, diagnostics such as those used by
\cite{ODea2008,Quillen2008}, such as the Spitzer IRAC
4.5$\mu$m/3.6$\mu$m colour and $\ohb$, may be used to discriminate
between the presence of an AGN and star formation.

We compiled the optical line emission ratio, $\ohb$, for {\cen}
available from \cite{Lewis2003} ($<0.4$) and \cite{Farage2010}
($\sim1$).  The two estimates are not necessarily contradictory since
these are calculated over different apertures. Neither of the ratios
indicates a presence of a dominant AGN contribution to the
far-infrared emission. In addition to the low $\ohb$, the IRAC
4.5$\mu$m/3.6$\mu$m colour ratio ($=0.58$) is consistent with a
passive stellar population. The absence of the IR {\oiii}~88{\mm}
spectral feature is another indication of negligible AGN
photoionization \citep{Spinoglio2009}.

\subsection{Gas Kinematics}
\label{kin}

Observations of {\cen} reveal interesting velocity structures and
several studies have used data from different wavebands to study them
\citep[e.g.][]{Sparks1989,Sparks1997,deJong1990}. The most recent
independent investigations by \cite{Farage2010} and \cite{Canning2011}
[in prep.]  use observations of {\ha} at sub-arcsecond resolution to
map the kinematics of the gas.

We used the projected map made using `specProject'
(Sect.~\ref{specanalysis2}) and obtained a line fit for each
individual spaxel separately.  The upper panel of Fig.~\ref{kinematics}
shows the line-of-sight velocity~(left) and the line FWHM~(right) the
{\cii} emission line gas as inferred with {\herschel}.  The X- and
Y-axis are the offsets along right ascension and declination relative
to the radio core position corresponding to the 5~GHz VLBA~(very large
baseline array) core maximum \citep{Taylor2006}.  These maps have a
nominal scale of $\pp{3}$ per spaxel. The lower panel of
Fig.~\ref{kinematics} shows the line-of-sight velocity~(left) and the
line FWHM~(right) of the {\ha}+{\niiopt} emission \citep{Canning2011}.
The {\ha}+{\niiopt} IFU~(integral field unit) observations were
performed with the VIMOS~(VIsible MultiObject Spectrograph) instrument
mounted on Melipal, a 8~m telescope at Paranal Observatory.

The line-of-sight velocity of the {\cii} emission varies from
$-115$~km~s$^{-1}$ to $120$~km~s$^{-1}$. The gas is blueshifted north
of the center (represented by the radio core), and redshifted south of
the center. This is consistent with the velocity map derived from
{\ha}+{\niiopt} emission which shows a similar range in the gas
velocity and shows that there is gas receding in the south and
approaching in the north. These results are also in agreement with the
velocity maps presented by \cite{Farage2010} using the Wide Field
Spectrograph~(WiFeS) instrument that operates on the ANU~(Australian
National University) 2.3~m telescope at Siding Spring Observatory.
Fig.~\ref{cencii} shows that the spatial morphology of the {\cii}
emission closely follows that of the X-ray and optical spiral
filaments.  Additionally, the velocity maps show very similar gas
kinematics in the infrared and the optical. The velocity and emission
maps from {\cii} and {\ha} suggest a geometry with gas infalling on a
spiral trajectory starting east of the galaxy center. The trajectory
loops around the galaxy in an anti-clockwise direction, such that the
infalling gas in the south appears to be receding in the rest frame of
the galaxy; hence redshifted. The trajectory continues and loops
around the galaxy towards the center, such that the gas in the north
appears to be approaching in the rest frame of the galaxy; hence
blueshifted.

The velocity dispersion of the {\cii} gas ranges from
$200~$km~s$^{-1}$ to $600~$km~s$^{-1}$. This is slightly larger than
the velocity dispersion in the {\ha}+{\niiopt} gas from the VLT/VIMOS
observations [50~km~s$^{-1}$ to 400~km~s$^{-1}$] and also from the
WiFeS observations [60~km~s$^{-1}$ to 525~km~s$^{-1}$]. The larger
values for the FIR {\cii} data than the optical {\ha} data is very
likely due to beam smearing, since the resolution of the Herschel
spectral data~($\pp{11}$) is a lot coarser than that of the optical
data~( $\pp{0.67}/$pix for the VLT and $\pp{0.5}/$pix for WiFeS)

The morphological and kinematical correlation between the far-infrared
forbidden line coolant, {\cii}, and the optical line filaments is a
key result of this work. This correlation has a profound implication,
namely that the optical hydrogen recombination line, {\ha}
\citep{Fabian1982,Crawford2005,Farage2010,Canning2011}, the optical
forbidden lines, {\niiopt} \citep{Dopita2010,Farage2010}, the soft
X-ray filaments \citep{Crawford2005} and the far-infrared {\cii} line
all have the same energy source.

\section{Modeling the interstellar medium of {\cen}}
\label{pdr}

To understand the complete picture giving rise to emission in {\cen}
at different wavebands ranging from X-ray to radio, we performed a
detailed investigation of the physical parameters of the interstellar
medium of the BCG.

The close spatial correspondence between the {\cii} maps, along with
the NW-SE elongated dust emission, and the optical {\ha} maps strongly
favours a common heating mechanism for the gas.  The excitation source
that leads to the optical line-emission and far-infrared coolants
remains to be investigated.  The Herschel observations presented in
this paper shed light on a key ingredient of this developing
picture. The positive detection of the most luminous far-infrared
coolant of the ISM, {\cii}, along with {\oi} and {\nii}, dictates the
presence of photodissociation regions. In this section we study the
source(s) responsible for heating the photodissociation
regions~(PDRs).

PDRs form a sharp interface with adjoining H{\sc ii}/H{\sc i} and
H{\sc ii} regions. Photons from nearby young stars (O- and B-type)
with energies $>13.6$~eV ionize the hydrogen in the molecular gas
giving rise to an H{\sc ii} region. This is followed by a PDR where
FUV photons with energies $6~$eV $< h\nu < 13.6$~eV are absorbed by
dust grains which re-radiate energy in the form of FIR continuum.
About 0.1~\% to 1~\% of the incident energy of the FUV photons is
converted into photoelectrons with energies $\leq 1~$eV which heat the
gas via elastic collisions. With CLOUDY we attempted to model an
integrated H{\sc ii}/PDR cloud, where the output emission originating
from the cloud also contains contribution from layers beneath the PDR
surface.

\subsection{Setting up the CLOUDY simulations}
\label{cloudysimulations}

\begin{table}
  \centering
  \caption{\small Observational constraints. The first column is the quantity, 
    the second column is the measured flux and the third column is the ratio 
    of the measured flux to the {\cii} 157.74{\mm} flux.}
  \label{pdr-constraints}
  \begin{tabular}{| c | c | c |}
    \hline
    Quantity   &  Value  ($10^{-15}~$erg~s$^{-1}$~cm$^{-2}$)  & F($\lambda$)/$F_{\textrm{C{\sc ii}}}$\\
    \hline
    \hline \rul
    {\cii}$^{\st c}$ 157{\mm}    & 174.7$\pm$3.1                                  &  1 \\ \rul
    $\ffir^{\st {a, c}}$               & (1.105$\pm$0.119)$\times10^{4}$ & 63.3$^{+17.3}_{-6.1}$  \\ \rul
    {\oi}$^{\st c}$ 63{\mm}    & 57.6$\pm7.7$                                 & 0.33$^{+0.07}_{-0.09}$ \\\rul
    {\nii}$^{\st c}$ 122{\mm} & 24.9$\pm1.7$                                 & 0.13$^{+0.01}_{-0.01}$ \\\rul
    {\ha}$^{\st b}$                  & 79.0$\pm0.9$                                 & 0.45$^{+0.05}_{-0.05}$ \\\rul
    {\si}$^{\st c}$                  & $< 4$                                 & $< 0.023$ \\\rul
    {\oiii}$^{\st c}$                  & $< 3$                                 & $< 0.017$ \\\rul
    {\oib}$^{\st c}$                  & $< 2$                                 & $< 0.011$ \\
    \hline
    \multicolumn{3}{p{8.5cm}}{ $^{\st a}$~Integrated FIR flux in the range (40--500)~{\mm}}\\
    \multicolumn{3}{p{8.5cm}}{$^{\st b}$~Extinction corrected {\ha} from \cite{Farage2010}}\\
    \multicolumn{3}{p{8.5cm}}{$^{\st {c}}$ This work. {\si}, {\oiii} and {\oib} are non-detections.}\\
  \end{tabular}
\end{table}

A detailed modeling of the various observed continuum and line
emission in {\cen} was conducted using the spectral synthesis code for
modeling the photodissociation regions, CLOUDY \citep[version 08.00
described by][]{Ferland1998}. CLOUDY models the clouds based on
conservation of energy through balancing the cooling and heating
rates. The simulations are self-consistent and include chemistry,
radiative-transfer and thermal balance. The basic ingredients of each
simulation comprised an old stellar population (OSP) and a young
stellar population (YSP), wherein the Starburst99 synthetic library
\citep{Leitherer1999,Vazquez2005} was used to model the spectra of
both the stellar populations. In addition, a background radiation was
included comprising the UVX cosmic background from radio to X-ray
\citep[e.g.][]{Ikeuchi1986,Vedel1994}, which also included the cosmic
microwave background radiation.

For the YSP, we used the same spectrum as used to derive the FUV star
formation rate in Sect.~\ref{sfr}, where the oldest stellar age was
set to $10^7$~yr and the star formation rate was considered to be
continuous and equal to $0.1~\mpy$. For the OSP, the oldest stellar
age was set to $10$~Gyr and the mass was fixed to $10^{11}~\ms$. Among
the specified model parameters were the normalizations for the OSP,
{\norm}, and the YSP, $G0$, in units of the \cite{Habing1968} field,
where one Habing is the interstellar radiation field of
$1.6\times10^{-3}$~erg~s$^{-1}$~cm$^{-2}$ in the energy range 6~eV and
13.6~eV. Another parameter was the total hydrogen density (atomic,
molecular, ionized and other hydrogen-bearing molecules) in units of
cm$^{-3}$.

\begin{table*}
  \centering
  \caption{\small The basic photodissociation region~(PDR) model parameters. The geometry is assumed to be plane-parallel. Columns: (1) parameter, (2) symbol, (3) the input range and (4) the most likely parameter values.}
  \label{pdr-model}
  \begin{tabular}{| c | c | c | c |}
    \hline
     Parameter   &  Symbol  & Input Range & Likely Values\\
     \hline
     \hline\rul
    Total Hydrogen Density (cm$^{-3}$)        &  $n$           & 10 to 10$^6$ & 50 to 100 \\\rul
    FUV Intensity Field  (Habing$^*$)                   &  $G0$         & 1 to 10$^6$ & 10 to 80  \\\rul
    Extra Heating (erg~s$^{-1}$~cm$^{-3}$)   &  $\hextra$ & $10^{-24}$ to $10^{-20}$ & $\sim 10^{-22}$\\\rul
    Normalization for the OSP$^*$ (10$^{-16}$~erg~s$^{-1}$~cm$^{-2}$~Hz$^{-1}$)&  {\norm} & 2.4 to 156.7 & 2.4 \\\rul
    Hydrogen Column Density (cm$^{-2}$)& {\hcol} & 10$^{19}$ to 10$^{23}$ & 10$^{21}$ \\\rul
    Metallicity  & $Z$ & 1 $-$ 4 & 2.5 \\
    Nitrogen abundance  (relative to $Z$) & {\nitrogen} & 2 $-$ 2.5 & 2 \\
    \hline 
    \multicolumn{4}{c}{$^{\st a}$~1~Habing $=1.6\times10^{-3}$~erg~s$^{-1}$~cm$^{-2}$, OSP: Old Stellar Population } \\
  \end{tabular}
\end{table*}

 \begin{figure*}
  \begin{minipage}{0.5\textwidth}
    \centering
    \includegraphics[width=\textwidth]{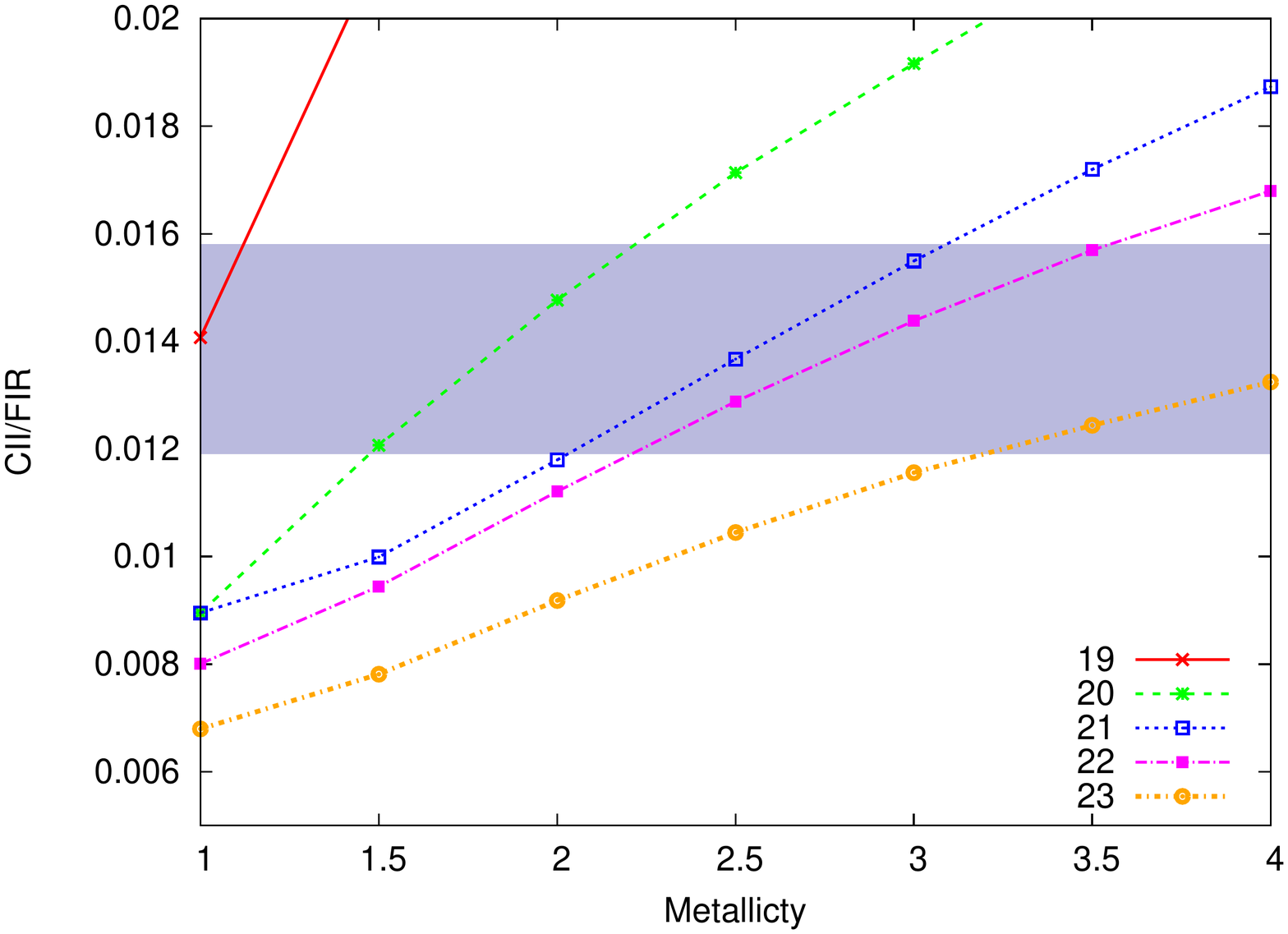}
  \end{minipage}%
  \begin{minipage}{0.5\textwidth}
    \centering
    \includegraphics[width=\textwidth]{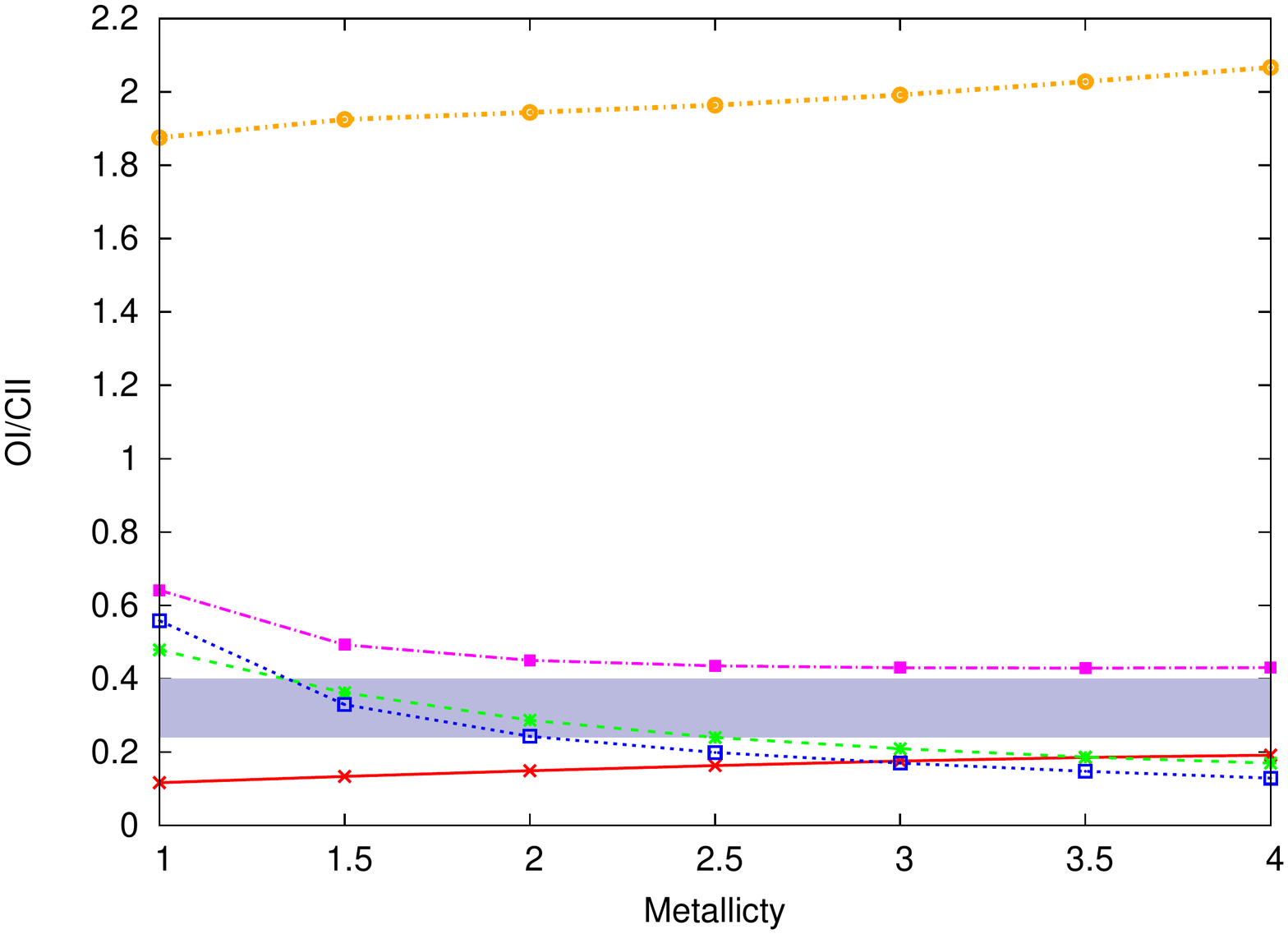}
  \end{minipage}\\
  \begin{minipage}{0.5\textwidth}
    \vspace*{-0.75cm}
    \centering
    \includegraphics[width=\textwidth]{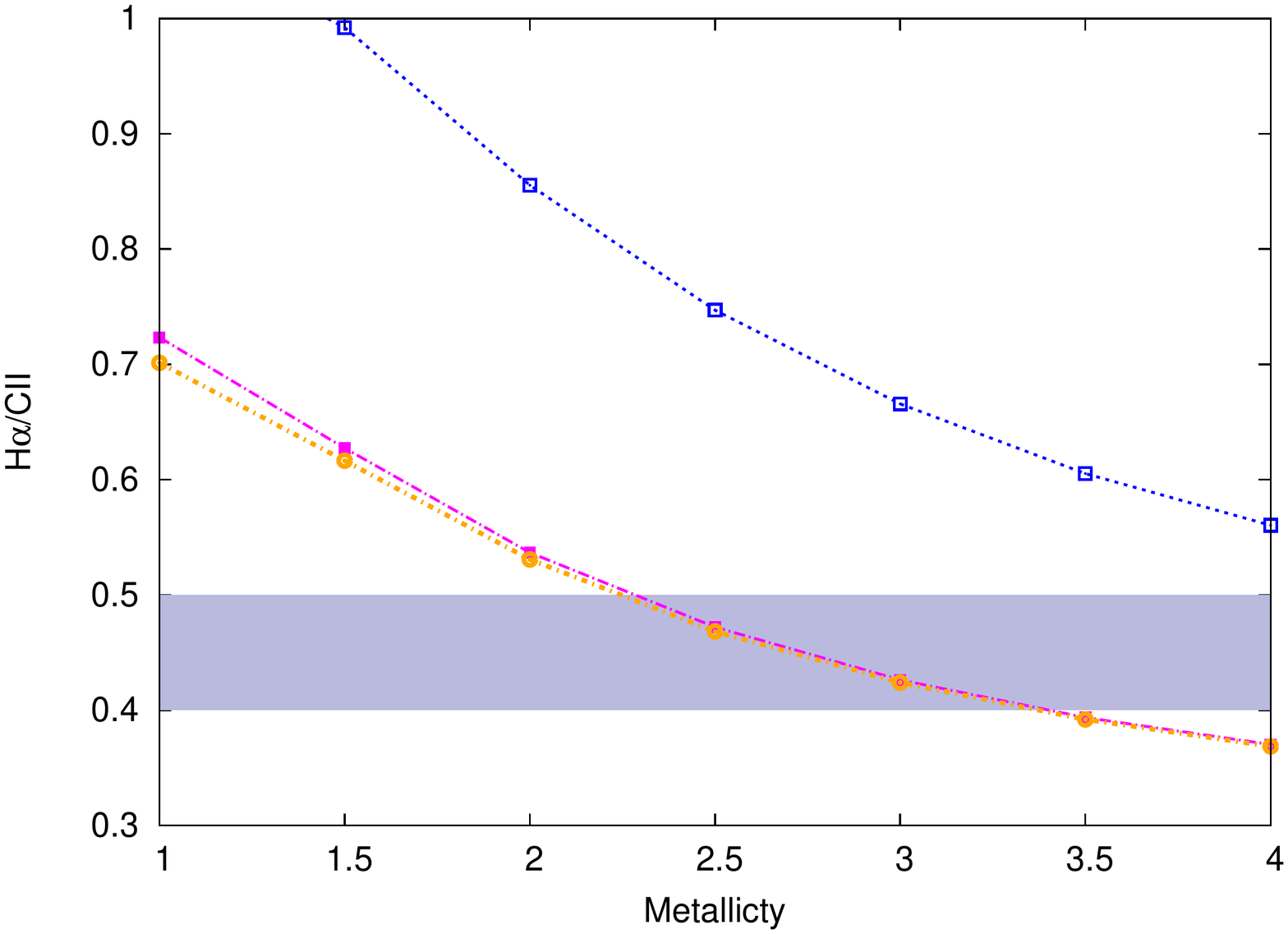}
  \end{minipage}%
  \begin{minipage}{0.5\textwidth}
    \vspace*{-0.75cm}
    \centering
    \includegraphics[width=\textwidth]{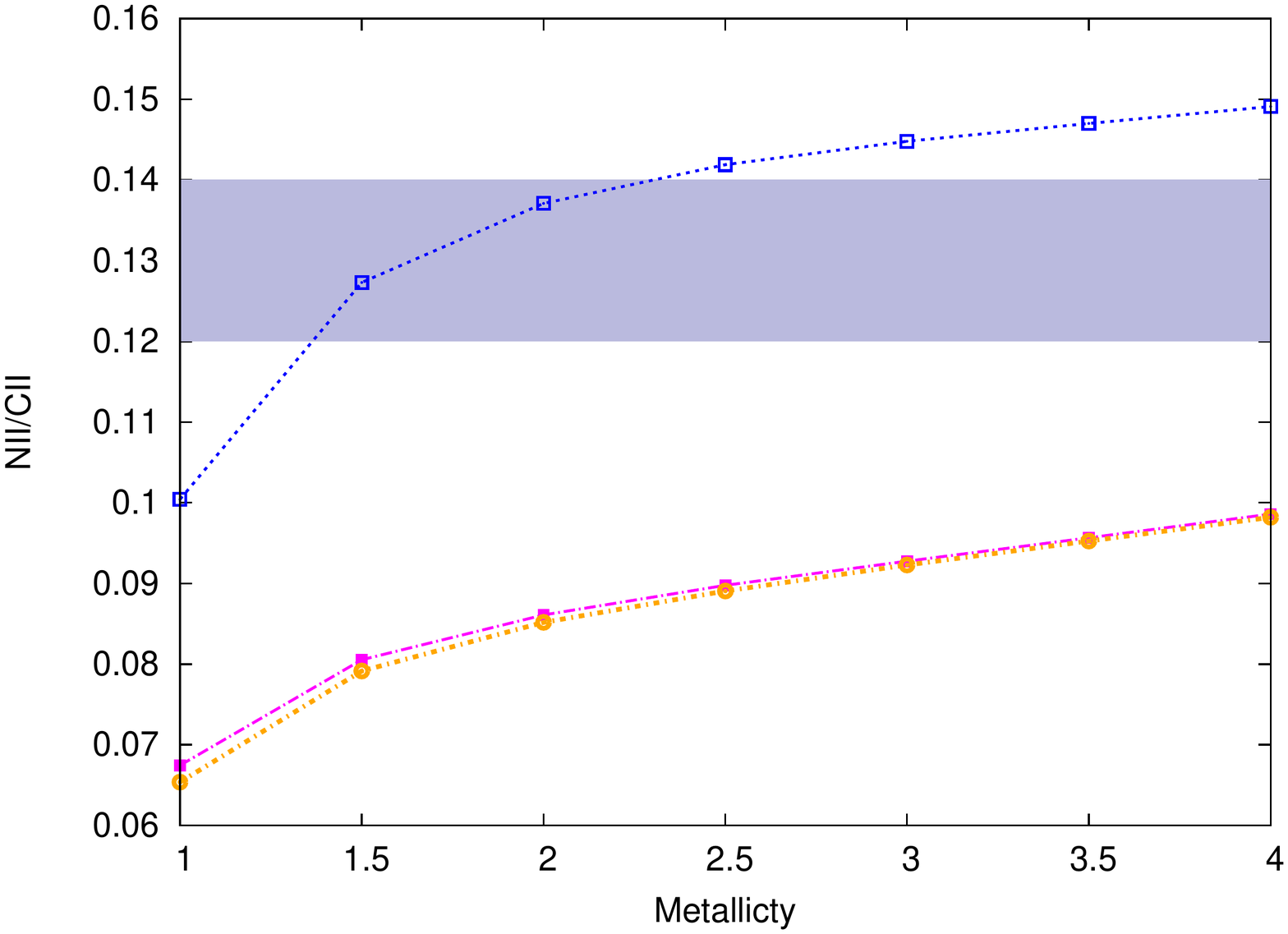}
  \end{minipage}
  \caption{\small The metallicity and penetration depth dependency of
    the line ratios, {\cii}/$\lfir$ (upper left), {\oi}/{\cii} (upper
    right), {\ha}/{\cii} (lower left) and {\nii}/{\cii} (lower
    right). The different lines correspond to the different
    penetration column densities, {\hcol} = 10$^{19}~$cm$^{-2}$ (red
    solid line), 10$^{20}~$cm$^{-2}$ (cyan dashed line),
    10$^{21}~$cm$^{-2}$ (dark blue dotted line), 10$^{22}~$cm$^{-2}$
    (magenta long-dashed dotted line) and 10$^{23}~$cm$^{-2}$ (orange
    short-dashed dotted line). The other parameters have been fixed to
    $G0 = 30$~Habing, $n$ = 100~cm$^{-3}$,
    $\hextra=10^{-22}$~erg~s$^{-1}$~cm$^{-3}$. The dust and nitrogen
    abundances are 1.2 times and 2 times the default ISM values. The
    shaded regions correspond to the upper and lower limit of the
    observed ratios.}
  \label{Zhtot}
\end{figure*}

This study and several other studies of {\cen} provide observational
constraints, which may be exploited to constrain the physical
properties of the ISM of the galaxy. The data explicitly used to
determine the most likely scenario are listed in
Table~\ref{pdr-constraints}. The errorbars on {\cii}/$\lfir$ were
estimated based on the minimum and maximum of the possible values
derived from $\lfir$ calculated with the 6 Herschel + MIPS 24{\mm}
data points and $\lfir$ calculated with the averaged data points in
case of duplicate measurements (see Sect.~\ref{dust}). The lower
errorbar on {\oi}/{\cii} was derived by integrating the flux over all
the spaxels with a SNR$>3$ for the {\cii} line and a SNR$>2$ for the
{\oi} line (because of the lower instrument sensitivity at the
wavelength of the {\oi} line emission and a generally weaker emission
relative to {\cii}). The upper errorbar on {\oi}/{\cii} was derived by
assigning the spaxels with no {\oi} detection but a positive {\cii}
detection a bare minimum flux equal to the sensitivity of the PACS
spectrometer in the third order times the chosen SNR, and likewise for
the {\cii} line. The errorbars on {\nii}/{\cii} and {\ha}/{\cii}
reflect the statistical uncertainties in the measurements.

In addition, \cite{Kaneda2005} found the IRS~({\spitzer} Infrared
Spectrograph) spectrum of {\cen} to show only one significant
polycyclic aromatic hydrocarbon~(PAH) feature at 12.7{\mm} with a flux
estimate of $14.9\times 10^{-15}$~erg~s$^{-1}$~cm$^{-2}$. PAH features
are very prominent in the vicinity of young massive stars and so are
considered as good tracers of star-forming sites. Absence of any
strong PAH features is in concert with the fact that both the FIR- and
FUV-derived star formation rates are low.  Also, even though the
dust-to-gas ratio in {\cen} is higher than the nominal range
(Sect.~\ref{gdmr}), PAHs owing to their small size are susceptible to
grain destruction via physical sputtering or thermal evaporation
\citep{Dwek1992}. While PAHs have an important effect on the chemistry
of PDRs \cite{Bakes1998}, due to the lack of any strong PAH features
in the IRS spectrum we did not consider PAH grains in our simulations.

The geometry was chosen to be plane-parallel by making the inner
radius much larger than the thickness of the cloud. The incident
radiation from the different stellar populations was input in units of
flux (erg~s$^{-1}$~cm$^{-2}$). The simulated output fluxes were
determined relative to the {\cii} flux and compared to the
observations. As a check on the absolute flux levels, we determined
the luminosity of the simulated {\cii} 157.74{\mm} line by multiplying
the flux with the surface area of the emitting cloud and scaling it to
the distance of the Earth. Although the radius of the cloud can be
judged from the size of the sphere encompassing the {\cii} emission
shown in Fig.~\ref{kinematics} as $R \sim \pp{17}$~($\sim 3.5~$kpc)
(also see Sect.\ref{lines}), the {\cii} flux is not expected to fill
up this whole sphere. The volume filling factor can be determined by
equating the simulated and observed fluxes. Note that the PDR model
thus constructed is oversimplified in that a single uniform PDR cloud
has been assumed. In reality, one expects a distribution of PDR clouds
with different and inhomogeneous incident and emergent
radiations. However, as shown below, a simple model such as described
here is sufficient to study some of the basic physical properties of
the ISM in {\cen}.

X-ray observations of the Centaurus cluster show supersolar abundances
in the inner 25~kpc, where the metallicity is $\sim 2$
\citep{Graham2006}. Hence we increased the elemental abundances, $Z$,
to two to three times the default ISM values. Furthermore, owing to a
small gas-to-dust mass ratio derived in Sect.~\ref{gdmr}, we increased
the dust abundance to 1.2 times the default ISM value leading to a
commensurate decrease in the gas-to-dust mass ratio from $\sim 150$ to
$\sim 70$.  Another important parameter is the penetration depth of
the incident radiation into the PDR cloud. The outer radius of the
cloud was specified in terms of the total hydrogen column density,
{\hcol}, at the edge of the cloud (the PDR surface). While the
observed ratio of {\nii} to {\cii} demands relatively low {\hcol}, the
observed ratio of {\ha} to {\cii} demands high {\hcol}. We
investigated a range of {\hcol} and $Z$ for the most likely values to
explain the observed ratios. The dependency of the ratios on {\hcol}
and $Z$ is shown in Fig.~\ref{Zhtot}. Based on the results of this
investigation, we fixed $Z$ to 2.5 and {\hcol} to $10^{21}$~cm$^{-2}$.
The various model parameters and their input values can be found in
Table~\ref{pdr-model}.

\subsection{Photoionization by stars}
\label{photoion}

\begin{figure}
  \centering
  \includegraphics[width=0.5\textwidth]{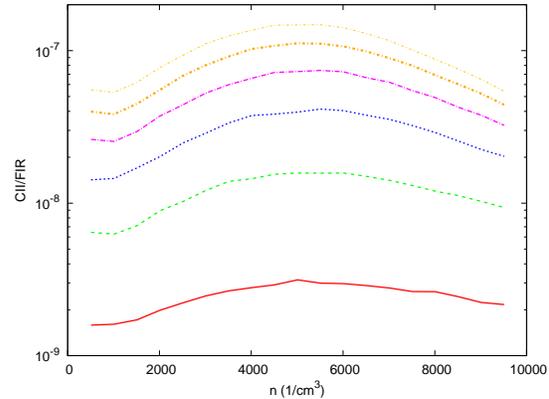}
  \caption{\small Simulations without any young stellar population
    produce low levels of {\cii} emission in comparison with the
    observations. As a result the {\cii}/$\lfir$ ratio is several
    orders of magnitude below the observed value. The different curves
    correspond to the different levels of {\norm}, the normalization
    for the old stellar population, increasing top to bottom from
    $2.4\times10^{-16}$~erg~s$^{-1}$~cm$^{-2}$~Hz$^{-1}$ (yellow
    short-dashed dotted line) to
    $156.7\times10^{-16}$~erg~s$^{-1}$~cm$^{-2}$~Hz$^{-1}$ (red solid
    line).}
  \label{noysp}
\end{figure}

The aim of conducting these simulations was to understand the key
physical ingredients behind the extended {\cii} and {\ha}
filaments. Is a young stellar population important? Can the old and
young stellar populations alone reproduce the observed ratios listed
in Table~\ref{pdr-constraints}? Do the observational constraints
require an additional source of heating? In the AGN-heating regulated
feedback framework, the AGN outflows return a fraction of the accreted
energy back to the intracluster gas. The radio source can heat the gas
either via energetic cosmic ray particles or inflating cavities. The
other scenario entailing additional heating revolves around
merger-induced shocks in the ISM of the BCG (see
Sect.~\ref{conclusions}).  

The young stellar population is important in terms of providing FUV
photons capable of ionizing carbon to produce the observed {\cii}
emission. Without any young stars, the {\cii}/$\lfir$ ratio can not be
reproduced. This is shown in Fig.~\ref{noysp}, where different curves
represent different values of {\norm}. The old stellar population
produces optical and UV photons which heat up the dust but does not
affect the gas much. Hence increasing {\norm} has an effect of
increasing $\lfir$ but not the intensity of the gas line emissions,
which decreases the {\cii}/$\lfir$ ratio. Though elementary, this
simple exercise underlines the importance of having both the stellar
populations. Furthermore, the observed {\cii} to $\lfir$ ratio ranges
from 0.1~\% in normal and starburst galaxies to 1~\% in dwarf
irregular galaxies \citep{Kaufman1999,Luhman2003}.
\cite{Edge2010a,Edge2010b} studied the {\cii}/$\lfir$ ratio in two
cool-core BCGs and found it to be equal to $\sim 0.4\%$ for Abell~1068
and $\sim 1.2\%$ for Abell~2597. In comparison, the observed ratio for
{\cen} is $\sim1.6$~\%. Thus the ratio is higher than normal and
requires minimum amount of dust heating. While such a large ratio may
be reconciled with the PDRs in low-metallicity galaxies with low
dust-to-gas mass ratios, {\cen} has a very high dust-to-gas mass
ratio. In order to reproduce the observed {\cii} to $\lfir$ ratio, we
fixed the {\norm} to have a bare minimum value of
$2.4\times10^{-16}$~erg~s$^{-1}$~cm$^{-2}$~Hz$^{-1}$. A lower value of
{\norm} does not produce any discernible effects on the output ratios.

Following this, we did simulations based on a model containing both
the old and young stellar populations, and probed $n$ and $G0$ space
ranging from (10 to $10^6$)~cm$^{-3}$ and (10 to $10^6$)~Habing.  The
left panel of Fig.~\ref{simulations} shows the contours of the
simulated ratios {\oi}/{\cii} and {\cii}/$\lfir$ corresponding to the
lower and upper limit given in Table~\ref{pdr-constraints} (where
reproducible). We find that the {\cii}/$\lfir$ ratio (dotted blue
line) saturates at an upper limit of about 0.014, where as the
observed ratio is 0.0147. Also, with this model the {\ha}/{\cii} ratio
is greater than 1 for all values of $n$ and $G0$, where as the
observed ratio is close to 0.45. Since neither the observed
{\cii}/$\lfir$ nor the {\ha}/{\cii} can be reproduced, we conclude the
model containing only old and young stellar populations as the source
of heating the PDRs is insufficient to explain the observations of
{\cen}.

\subsection{Extra heating}
\label{extra}

The gas in PDRs is mostly heated via the dust through photoelectric
heating. In order to boost the {\cii} flux independent of the FIR dust
emission, a source of heating is required that may directly input
energy to the gas. Thus we included an extra heating term in CLOUDY
using the parameter ``hextra'', which specifies a volume-heating rate
(erg~s$^{-1}$~cm$^{-3}$). The source of this extra heating is
unspecified and so may be attributed to either shock heating or cosmic
ray heating due to the radio source.  We investigated models
containing the OSP and YSP and additional heating ranging from
$10^{-24}$~erg~s$^{-1}$~cm$^{-3}$ to
$10^{-20}$~erg~s$^{-1}$~cm$^{-3}$. The term ``hextra'' acts to raise
the temperature of both the gas and the dust independent of each
other. The standard behaviour of {\cii} is such that for a given $G0$
the {\cii} intensity increases linearly up to the critical density ($n
\sim 3000$~cm$^{-3}$ for the {\cii} forbidden line) beyond which
collisional de-excitation occurs as often as radiation. Hence the line
intensity above the critical density is not very sensitive to the
density and increases with it only very slowly. The extra heating acts
to boost the {\cii} intensity at low densities all the way up to $n
\sim 10^{4}$~ cm$^{-3}$ after which it decreases due to overly
suppressed radiative de-excitation. The extra heating also increases
the dust emission; however the increase is roughly independent of the
density. The over all effect of the heating term on the {\cii}/$\lfir$
ratio is such that the ratio increases for low densities, where the
increase in {\cii} is larger than the increase in $\lfir$.

\begin{figure*}
  \centering
  \begin{minipage}{0.33\textwidth}
    \centering
    \includegraphics[width=0.9\textwidth]{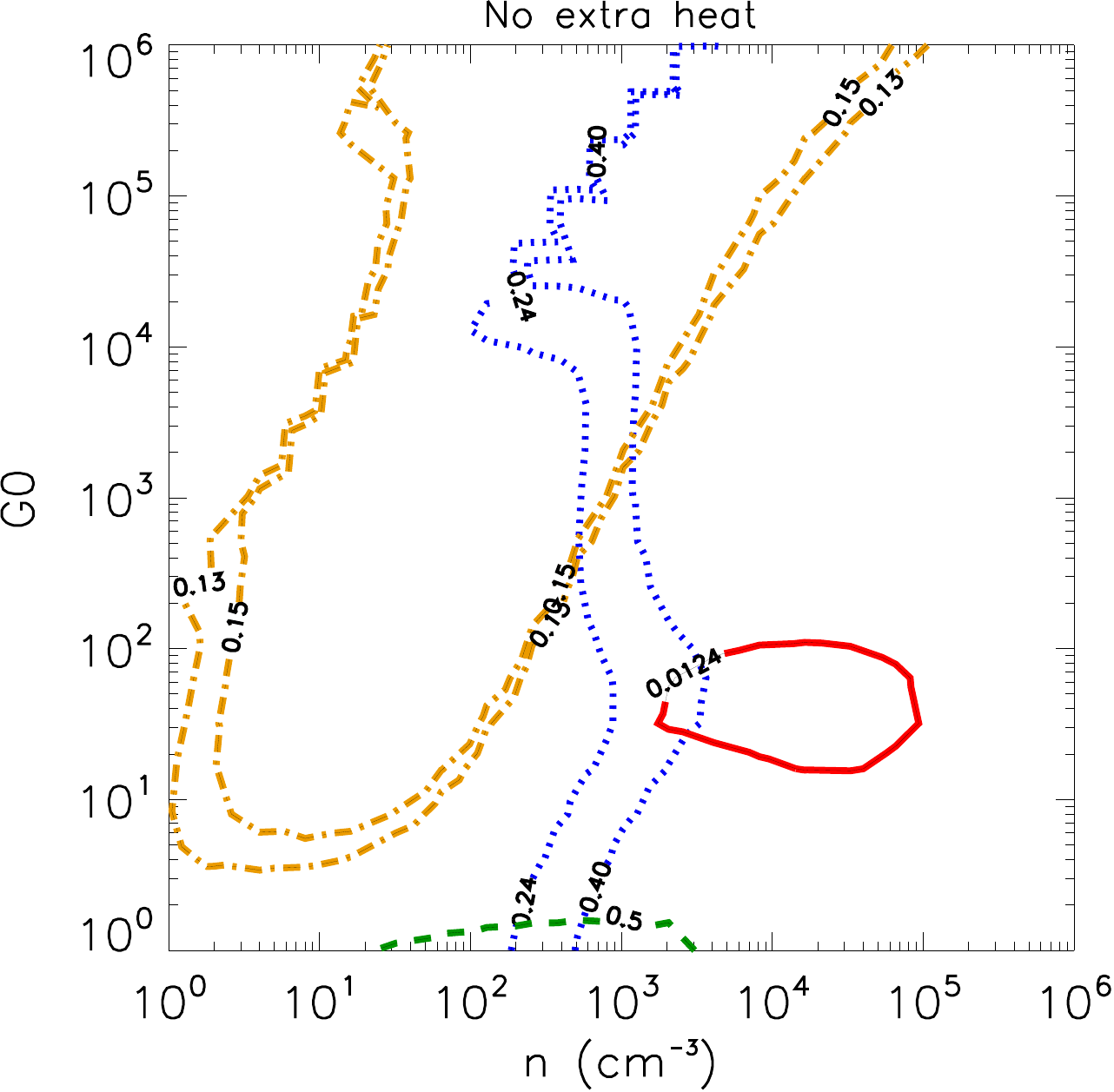}
  \end{minipage}%
  \begin{minipage}{0.33\textwidth}
    \centering
    \includegraphics[width=0.9\textwidth]{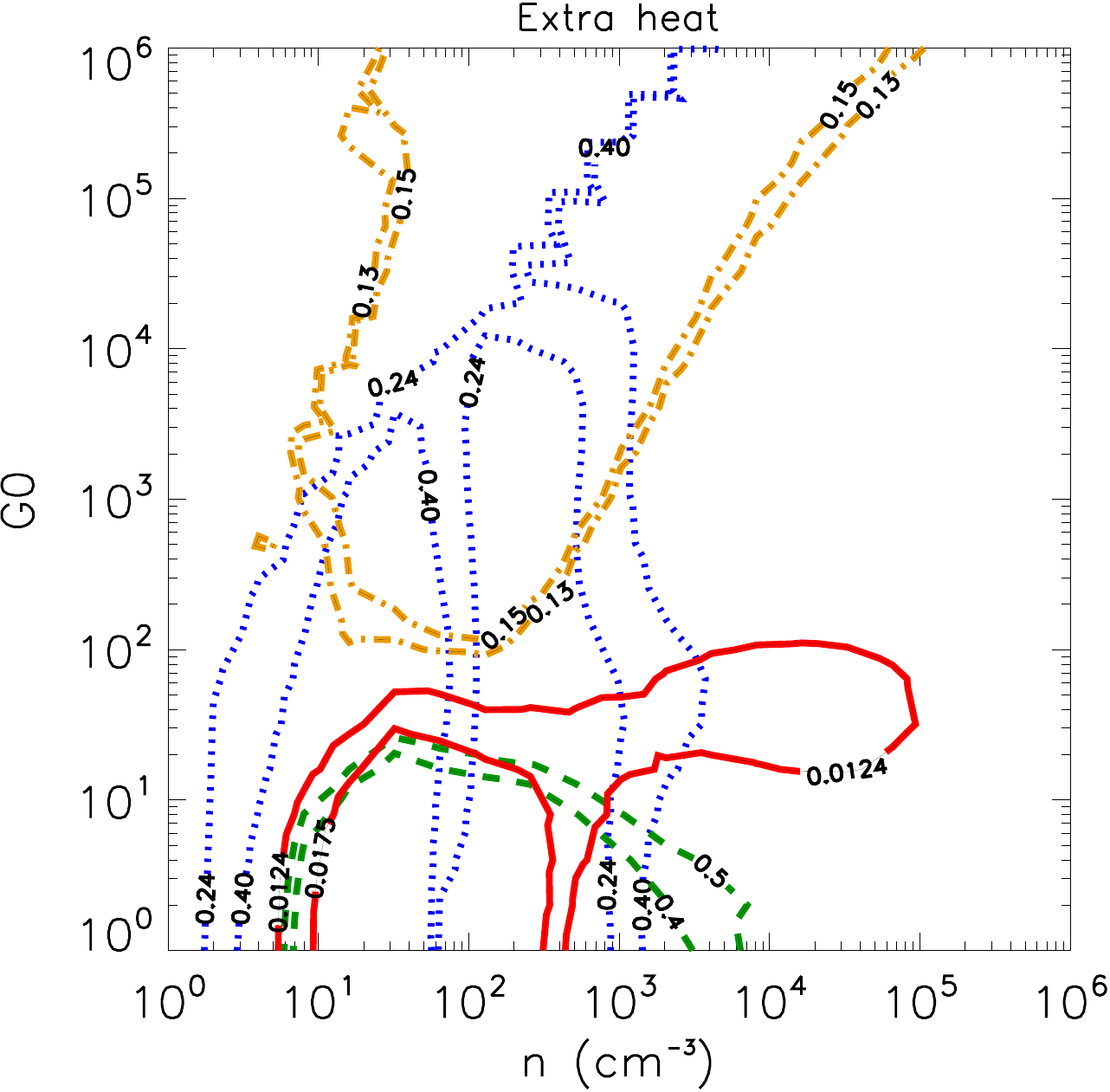}
  \end{minipage}%
  \begin{minipage}{0.33\textwidth}
    \centering
    \includegraphics[width=0.9\textwidth]{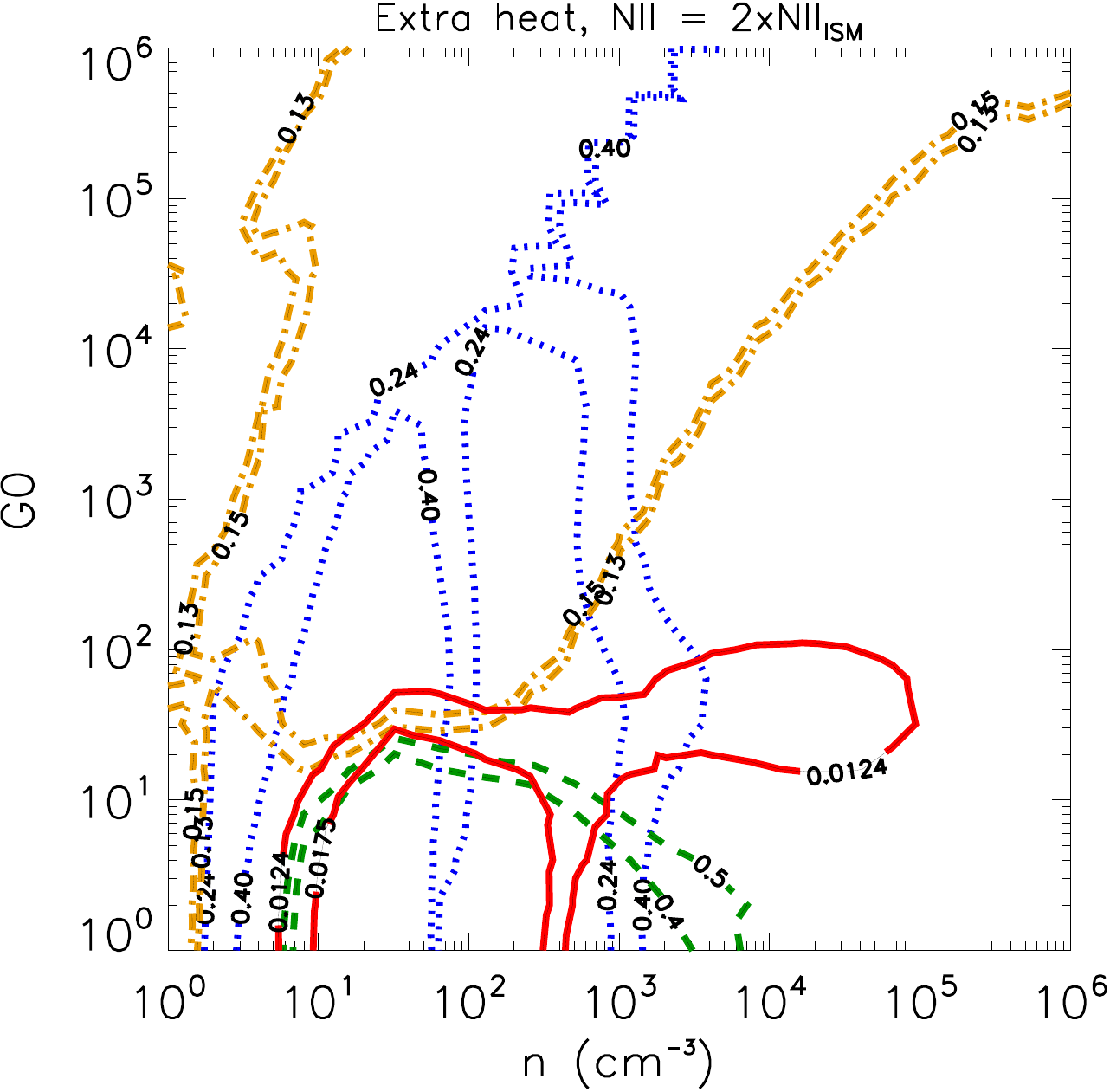}
  \end{minipage}%
  \caption{\small Simulations containing both old and young stellar
    populations. The dust abundance is set to 1.2 times the ISM value
    and the metallicity is set to 2.5 times the ISM value. Shown are
    the contours corresponding to the lower and upper observed limits
    (where reproducible) for {\cii}/$\lfir$ (solid red line),
    {\oi}/{\cii} (dotted blue line), {\ha}/{\cii} (dashed green line)
    and {\nii}/{\cii}, (dashed-dotted mustard line). {\it Left}:
    Heating by stars only. {\it Middle}: Heating by stars and an
    additional source, such as cosmic-rays, shocks etc. {\it Right}:
    Nitrogen overabundance of a factor of two.}
  \label{simulations}
\end{figure*}

The middle panel of Fig.~\ref{simulations} shows the contours of
{\cii}/$\lfir$ (solid red), {\oi}/{\cii} (dotted blue) and
{\ha}/{\cii} (dashed green) for
$\hextra=10^{-22}$~erg~s$^{-1}$~cm$^{-3}$. Of the different levels of
$\hextra$ investigated, this was the optimal value leading to most
coherent results. For example,
$\hextra=10^{-21}$~erg~s$^{-1}$~cm$^{-3}$ fails to reproduce the
observed {\oi}/{\cii} ratio and
$\hextra=10^{-23}$~erg~s$^{-1}$~cm$^{-3}$ just barely touches the
observed {\cii}/$\lfir$ ratio. We found that while $\hextra$ could
reproduce the right levels of {\cii}/$\lfir$, {\ha}/{\cii} and
{\oi}/{\cii}, the {\nii}/{\cii} ratio was too low compared to the
observations. The FIR {\nii} line emission (the first ionization
potential of nitrogen is 14.5~eV) arises from warm ionized medium
only, where as {\cii} emission can arise from both neutral (PDRs) and
ionized media. The fraction of {\cii} emission arising from ionized
media is highly uncertain, ranging from 25~\% up to more than 50~\%
\citep[e.g.][]{Heiles1994,Oberst2006,Aannestad2003}. However, since
our model includes a PDR adjacent to an H{\sc ii} region, the total
{\cii} emission includes contributions from both, neutral as well as
ionized media, and no correction factor is needed to estimate the
fraction of {\cii} emission arising from the PDR alone. The right
panel of Fig.~\ref{simulations} corresponds to the same setting as for
the middle panel, only the nitrogen abundance, {\nitrogen}, has been
increased by a factor of two over the ISM value.  The contours
corresponding to the {\nii}/{\cii} ratio (dotted-dashed mustard) now
overlap with the contours of the {\cii}/$\lfir$, {\oi}/{\cii} and
{\ha}/{\cii} ratios. A nitrogen overabundance from the PDR analysis is
consistent with the enhancement in nitrogen abundance required based
on the optical line ratios \citep{Farage2010} and X-ray spectral
fitting \citep{Sanders2008}. The fact that the analyses of the {\cen}
observations in X-ray, optical and far infrared wavebands, all require
a nitrogen overabundance and by roughly the same factor is reassuring
and lends support to the above PDR model.

The emanating {\cii} flux from the cloud is about
\mbox{$1\times10^{-3}$~erg~s$^{-1}$~cm$^2$}. The size of the cloud
that results in the observed {\cii} flux is such that,
\mbox{$F_{\textrm{C{\sc ii}, obs}} = F_{\textrm{C{\sc ii}, mod}}
  \times (r/\Dl)^2$}, where $F_{\textrm{C{\sc ii}, obs}}$ is the
observed {\cii} line flux and $F_{\textrm{C{\sc ii}, mod}}$ is the
modelled {\cii} line flux, $r$ is the ``effective radius'' of the
cloud and $\Dl$ is the luminosity distance to {\cen}.  The radius of
the cloud thus calculated is $r=0.5$~kpc. Considering that the {\cii}
flux extends over a $\sim 3.5$~kpc large region implies a volume
filling factor of $\sim 3\times10^{-3}$.

These PDR simulations help us to narrow down the most probable heating
scenario such that it simultaneously reproduces all the observed line
and continuum ratios. The most likely parameter values are given in
the last column of Table~\ref{pdr-model}. The right panel of
Fig.~\ref{simulations} is based on a model that contains
photoionization from old and young stellar populations, and an
additional source of heating. Note that photoionization from young
stars is still of paramount importance. A model containing only extra
heating and no young stars produces very little {\cii} because of
severe lack on FUV photons.  The model containing stars and
``extra-heating'' best explains the observations yielding a density,
$n$, in the range of about a few tens to hundred per cm$^3$ and a
radiation field, $G0$, in the range ten Habing to $\sim 80$ Habing.
This range of $G0$ corresponds to a star formation rate of (0.02 to
0.13)~$\mpy$ for a cloud of an effective radius of $0.5$~kpc. The SFR
thus estimated agrees very well with the limits derived from the FIR
and NUV/FUV observations (Sect.~\ref{sfr}).

\section{Discussion}
\label{conclusions}

The Centaurus cluster of galaxies can be classified as a strong
cool-core cluster based on its short gas cooling time ($< 0.5~$Gyr)
and a peaked X-ray surface brightness profile. The mass condensation
rate derived from the far-infrared and UV observations is small ($\sim
0.1 \mpy$) as compared to that predicted based on the cooling-flow
model ($\sim 25~\mpy$). The X-ray core luminosity at 2.5~\%$\rvir$
corresponding to a region of radius 30~kpc is about
$1.3\times10^{43}$~erg~s$^{-1}$ \citep{Sanders2008}. This is the
amount of power (radiative losses) that needs to be fed back to the
ICM over that scale for the system to remain in equilibrium. One of
the potential feedback mechanisms is through the expanding radio
plasma that has dug out multiple cavities in the intracluster medium
of {\cen} at radii $\leq 6$~kpc. The power in these radio bubbles
assuming a buoyancy timescale is estimated to be about
$0.7\times10^{43}$~erg~s$^{-1}$ \citep{Rafferty2006}. Repetitive
inflation of such radio bubbles also creates sound waves which
generate a sound-wave power of about $0.5\times10^{43}$~erg~s$^{-1}$
\citep{Sanders2008}. These two mechanisms together suffice to balance
the X-ray radiative losses and can keep the central region of
Abell~3526 in quasi-equilibrium.

The optical and far-infrared line emissions clearly show an infalling
gas flow of spiral pattern. There are two possibilities for the origin
of the gas and dust. The first possibility is that this material
represents the cooling intracluster medium flowing toward the center
of the dark matter potential well (which in the case of Centaurus
coincides with the BCG). The fact that the soft X-ray emission
(0.5-1)~keV traces the spiral structure seen in {\ha} and {\cii} does
indeed suggest a link to the gas cooling out of the ICM.  The relative
increase in the X-ray surface brightness along the filaments is a
natural outcome of this scenario.  \cite{Fabian2011} examined the
Chandra X-ray emission from the Northern filament in the Perseus
cluster and provided another outlook on the excitation mechanism of
the filaments. Their results indicate that some of the soft X-ray
emission may be due to charge exchange as the hot ionized gas
penetrates and mixes with cold gas. Although the cooling flow model
does not dictate a coherent directional gas flow to the center, such
as that seen in {\cen}, the spiral morphology may be a natural
consequence of the global velocity field of the gas surrounding in
{\cen}. This may be the first direct mapping of a cooling flow in a
galaxy cluster.

The second possibility is that the filaments and the dust lane in
{\cen} have a galaxy infall origin. This is based on a merger
hypothesis proposed by \cite{Sparks1989} and more recently by
\cite{Farage2010}. Under this hypothesis, the BCG is undergoing a
minor merger with an infalling neighbouring dust-rich galaxy.  The
presence of two nuclei in HST imaging of the center of {\cen}
\citep{Laine2003} lends some support to the merger hypothesis,
although the nature of the second nucleus is not well established. A
merger hypothesis also explains the prominent dust lane seen with the
{\it HST}. While \cite{Sparks1989} favour heat transfer via conduction
from the hot X-ray emitting ICM to the cool dust-rich infalling gas,
\cite{Farage2010} favour shock heating generated through the
dissipation of the orbital energy of the infalling cloud.  Assuming a
shock luminosity of $2\times10^{42}$~erg~s$^{-1}$ \citep{Farage2010},
the shock volume luminosity over a region of radius 0.5~kpc is
$1.3\times10^{-22}$~erg~s$^{-1}$~cm$^{-3}$. This is in good agreement
with the level of the extra heating input to the simulations above.

Irrespective of the origin of the cold gas and dust, the detailed PDR
modeling of the observed quantities arising from this flow (described
in Sect.~\ref{pdr}) shows photoionization from stars is an important
requisite. However, there is an additional source of energy input
required to reproduce the observables. This additional source of
heating could either represent the energy deposited by the AGN through
the methods described above or shock heating due to a
minor-merger. The additional heating amounts to $< 10~\%$ of the X-ray
core luminosity and could be another contributing factor to the total
heating budget.

The non-radiative energy sources, in principle, could be input into
the simulations either in the form of extra heating \cite[shock
heating or reconnection diffusion, see][ for latter]{Fabian2011}, such
as that considered in these CLOUDY simulations, or cosmic
rays. \cite{Ferland2008} and \cite{Ferland2009} have shown that while
it is difficult to differentiate between the two processes, certain
optical lines could help discriminate between the two.  The extra
heating case in those studies has primarily been associated with
magneto-hydrodynamic~(MHD) waves such as those observed in the ISM of
our galaxy. The dissipation of MHD wave energy can heat the gas;
however, the heating simply adds to the thermal energy of the gas so
that the velocity distribution remains Maxwellian. On the other hand,
the cosmic ray heating by high-energy particles can both heat and
ionize the gas. Both the processes heat the gas but the cosmic ray
(ionizing-particle) case produces a population of first and second
ions by non-thermal collisional ionization. In the extra-heating case
these ions only occur when the gas is warm enough for collisional
ionization equilibrium to occur. This is the main distinction between
the two cases, extra heating and cosmic ray heating.

\cite{Ferland2009} studied the Horseshoe region of the Perseus galaxy
cluster and used the infrared and optical line intensities to
distinguish between extra heating and cosmic ray heating. They found
that both the heating cases match the observations to within a factor
of two for the majority of the lines. There are a few discriminant
lines, such as the optical emissions [He{\sc i}]~$\lambda~5876~\AA$
and [Ne {\sc iii}]~$\lambda~3869~\AA$ and the infrared emission [Ne
{\sc ii}]~$\lambda~12.81${\mm} which show a few orders of magnitude
difference and indicate that the cosmic ray may be a better agent for
heating and ionizing the gas \citep[also see][]{Donahue2011}. The
mid-infrared Spitzer IRS spectrum of the off-center regions of {\cen},
in fact, contains the [Ne {\sc ii}]~$\lambda~12.81${\mm} line
\citep{Johnstone2007}. This along with the detections of
ro-vibrational H$_2$ lines in off-nuclear regions in {\cen} and
NGC~1275 provided some of the motivation for the heating models
investigated by \cite{Ferland2009}.  However, previous studies did not
consider the lines observed with {\herschel}, which provide a
crucially important perspective on the physical model of the ISM.

Though our model contains additional energy explicitly in the form of
extra heating, we speculate the ionizing-particle model would also
have fitted the available data, as has been shown for Perseus. In
order to discern the two heating sources, deep optical observations
are needed to complement the Herschel data.

\section{Summary}
\label{summary}

We made far-infrared~(FIR) observations of the brightest galaxy,
{\cen}, of the Centaurus galaxy cluster with the {\herschel} telescope
to better understand the cooling and heating of the intracluster
medium.

\begin{itemize}
\item We have detected FIR coolants in {\cen}, which include extended
  {\cii}~157.74{\mm}, marginally extended {\oi}~63.18{\mm} and
  unresolved {\nii}~121.90{\mm} line emissions.
\item We have detected far-infrared dust emission from the BCG at
  70{\mm}, 100{\mm}, 160{\mm}, 250{\mm}, 350{\mm} and 500{\mm}. A
  spectral energy distribution~(SED) fitting of the dust emission
  reveals a high-mass cold component around $19~$K and a low-mass warm
  component around $60$~K.
\item Using the dust mass calculated from the SED fitting and the
  non-detection of CO \citep{ODea1994}, we derived an upper limit on
  the gas-to-dust mass ratio of 125. This makes {\cen} a galaxy with
  one of the lowest gas-to-dust mass ratios.
\item The star formation rate derived from the integrated FIR
  luminosity is about 0.13~$\mpy$. We derived similar upper limits
  from HST far-ultraviolet and GALEX near-ultraviolet observations.
\item The extended {\cii} emitting gas shows remarkably similar
  spatial morphology and kinematics as the optical {\ha} emitting and
  X-ray gas. This implies a common heating mechanism of the gas.
\item We envisage the FIR and optical emissions as arising from
  photo-dissociation regions~(PDRs) adjacent to ionized regions. From
  a detailed modeling of such an integrated PDR, we conclude that in
  addition to heating via stellar photoionization, an additional
  non-radiative heating is required. The most likely model yields a
  total hydrogen density in the range (50 to 100)~cm$^{-3}$ and a FUV
  intensity field in the range (10 to 80)~Habing.
\end{itemize}

\section*{Acknowledgments}
This work is based (in part) on observations made with Herschel, a
European Space Agency Cornerstone Mission with significant
participation by NASA. Support for this work was provided by NASA
through an award issued by JPL/Caltech.  We would like to thank the
HSC and NHSC consortium for support with data reduction
pipelines. R.~Mittal is grateful to P.~Appleton for repeated help with
the Herschel analysis and thanks J.~T.~Whelan and D.~Merritt for
comments and discussions. B.~McNamara and H.~Russell acknowledge
generous financial support from the Canadian Space Agency Space
Science Enhancement Program. N. Hatch thanks STFC and the University
of Nottingham Anne McLaren Fellowship for support. G.~Tremblay
acknowledges support from the New York Space Grant Consortium. This
research has made use of the NASA/IPAC Extragalactic Database (NED)
which is operated by the JPL/Caltech, under contract with NASA.
STSDAS is a product of the Space Telescope Science Institute, which is
operated by AURA for NASA.


{\small
\bibliographystyle{mn2e}
\bibliography{ref}}

\end{document}